\documentclass[sigconf]{acmart}

\usepackage{verbatim}
\usepackage{enumitem}
\usepackage{graphicx}
\usepackage{cleveref}

\usepackage{xspace}

\usepackage{xcolor}
\usepackage{color, colortbl}
\definecolor{gainsboro}{rgb}{0.86, 0.86, 0.86}

\usepackage{multirow}
\usepackage{array}
\newcolumntype{P}[1]{>{\centering\arraybackslash}p{#1}}

\usepackage{pifont}

\usepackage{tikz}
\usetikzlibrary{positioning, arrows.meta, decorations.pathreplacing, fit, backgrounds, patterns}

\usepackage{pgfplots}
\pgfplotsset{width=\columnwidth,compat=1.9}
\usepackage{caption}
\usepackage{subcaption}

\newcommand{\etal}{\emph{et al.}\xspace}
\newcommand{\cf}{cf.\xspace}

\newcommand{\cmark}{\ding{51}}
\newcommand{\xmark}{--}

\newcommand{\step}[1]{Step~\circled{#1}\xspace}
\newcommand*\circled[1]{\tikz[baseline=(char.base)]{\node[shape=circle,draw, solid,inner sep=0.5pt] (char) {#1};}}
\newcommand{\stepCircled}[1]{\circled{#1}\xspace}

\newcommand{\lib}{\textsc{Keyfort}\xspace}

\newcommand{\party}{\ensuremath{P}\xspace}

\newcommand{\SM}{\ensuremath{SM}\xspace}
\newcommand{\SMSrc}{\ensuremath{SM_S}\xspace}
\newcommand{\SMDst}{\ensuremath{SM_D}\xspace}

\newcommand{\E}{\ensuremath{E}\xspace}
\newcommand{\ESrc}{\ensuremath{E_S}\xspace}
\newcommand{\EDst}{\ensuremath{E_D}\xspace}

\newcommand{\StateSrc}{\ensuremath{D_S}\xspace}
\newcommand{\StateDst}{\ensuremath{D_D}\xspace}

\newcommand{\pkSrc}{\ensuremath{pk_S}\xspace}
\newcommand{\pkDst}{\ensuremath{pk_D}\xspace}

\newcommand{\ver}{\ensuremath{v}\xspace}
\newcommand{\vSrc}{\ensuremath{v_S}\xspace}
\newcommand{\vDst}{\ensuremath{v_D}\xspace}
\newcommand{\vlate}{\ensuremath{v_{latest}}\xspace}

\newcommand{\eid}{\ensuremath{eid}\xspace}
\newcommand{\eidSrc}{\ensuremath{eid_S}\xspace}
\newcommand{\eidDst}{\ensuremath{eid_D}\xspace}

\newcommand{\ID}{\ensuremath{ID}\xspace}
\newcommand{\IDSrc}{\ensuremath{ID_S}\xspace}
\newcommand{\IDDst}{\ensuremath{ID_D}\xspace}

\newcommand{\meas}{\ensuremath{m}\xspace}
\newcommand{\measDst}{\ensuremath{m_D}\xspace}
\newcommand{\measSrc}{\ensuremath{m_S}\xspace}

\newcommand{\seed}{\ensuremath{s}\xspace}
\newcommand{\key}{\ensuremath{k}\xspace}

\newcommand{\bin}{\ensuremath{b}\xspace}
\newcommand{\binSrc}{\ensuremath{b_S}\xspace}
\newcommand{\binDst}{\ensuremath{b_D}\xspace}

\newcommand{\enclTime}{\ensuremath{t_E}\xspace}
\newcommand{\entryTime}{\ensuremath{t_E^{entry}}\xspace}

\newcommand{\resumeok}{\ensuremath{resume\_ok}\xspace}

\newcommand{\mtime}{\texttt{mtime}\xspace}

\newcommand{\subheading}[1]{\vspace{0.25 em}\noindent \textbf{#1}}

\renewcommand\footnotetextcopyrightpermission[1]{}

\author{Annika Wilde}
\email{annika.wilde@rub.de}
\affiliation{%
	\institution{Ruhr University Bochum}
	\country{Germany}}

\author{Samira Briongos}
\email{samirabriongos@gmail.com}
\affiliation{%
	\institution{NEC Laboratories Europe}
	\country{Germany}}

\author{Claudio Soriente}
\email{csoriente@gmv.com}
\affiliation{%
	\institution{GMV Spain}
	\country{Spain}}

\author{Ghassan Karame}
\email{ghassan@karame.org}
\affiliation{%
	\institution{Ruhr University Bochum}
	\country{Germany}}

\begin{document}

\title{On Securing the Software Development Lifecycle in IoT RISC-V Trusted Execution Environments}

\begin{abstract}
	RISC-V–based Trusted Execution Environments (TEEs) are gaining traction in the automotive and IoT sectors as a foundation for protecting sensitive computations. However, the supporting infrastructure around these TEEs remains immature. In particular, mechanisms for secure enclave updates and migrations---essential for complete enclave lifecycle management---are largely absent from the evolving RISC-V ecosystem.
	
	In this paper, we address this limitation by introducing a novel toolkit that enables RISC-V TEEs to support critical aspects of the software development lifecycle. Our toolkit provides broad compatibility with existing and emerging RISC-V TEE implementations (e.g., Keystone and CURE), which are particularly promising for integration in the automotive industry. It extends the Security Monitor (SM)---the trusted firmware layer of RISC-V TEEs---with three modular extensions that enable secure enclave update, secure migration, state continuity, and trusted time.
	
	Our implementation demonstrates that the toolkit requires only minimal interface adaptation to accommodate TEE-specific naming conventions. Our evaluation results confirm that our proposal introduces negligible performance overhead: our state continuity solution incurs less than 1.5\% overhead, and enclave downtime remains as low as 0.8\% for realistic applications with a 1 KB state, which conforms with the requirements of most IoT and automotive applications.
\end{abstract}

\maketitle
\pagestyle{plain}

\section{Introduction}

Emerging technologies are being rapidly adopted across multiple industries, with the automotive and IoT sectors at the forefront of leveraging them to enhance consumer experiences. In the automotive domain, manufacturers are increasingly transitioning toward zonal architectures, consolidating electronic control units (ECUs) into fewer, more powerful nodes. This trend amplifies the need to protect sensitive user data and workloads.

Trusted Execution Environments (TEEs) are a cornerstone of secure and confidential computing, enabling sensitive workloads to run within isolated environments, or enclaves. In automotive applications, TEEs can securely process payments, such as tolls or parking fees, and isolate critical components within networks of interconnected ECUs~\cite{blog:embitel/automotive/TEEvsHSM,blog:embitel/automotive/TEEinECU}. According to the GlobalPlatform Automotive Task Force, over 100 million TEEs were actively deployed in vehicles as of 2023~\cite{globalplatformAutomotive}. Examples include the Snapdragon Cockpit Platform, built on automotive Snapdragon SoCs leveraging the Qualcomm Trusted Execution Environment~\cite{psacertifiedSnapdragonSA61xxP}, which powers the Mercedes-Benz User Experience~\cite{qualcomm2025snapdragon,qualcommMercedesBenz}, and Toyota Lexus vehicles equipped with Renesas R-Car SoCs for enhanced user experience~\cite{renesas2025toyota} running Trustonic’s Kinibi TEE OS~\cite{trustonicRenesasConsortium}.

Amid rising geopolitical tensions and the drive for technological independence, RISC-V---a modular, open standard instruction set architecture (ISA)---has emerged as a compelling foundation for next-generation automotive systems. Its flexibility, scalability, and security make it well-suited for building adaptable platforms that can integrate TEEs. Major automotive suppliers are preparing for RISC-V adoption: Mobileye’s EyeQ silicon, already embedded in over 200 million vehicles~\cite{mobileyestats}, is expected to transition the majority of its ADAS systems to RISC-V by 2030~\cite{riscvForAutomotiveAI}, and Infineon, the global market leader for automotive microcontrollers with a 28.5\% market share in 2023, announced a new RISC-V-based automotive MCU family in 2025~\cite{infineonRiscvForAutomotive}. \emph{These developments indicate that future automotive platforms are likely to incorporate RISC-V TEEs, making their adoption a tangible near-term reality.}

Unfortunately, no RISC-V–based TEE---and in fact, no commercial TEE design---offers secure, end-to-end software lifecycle support with features such as secure timekeeping and state-preserving software updates or enclave migrations. These capabilities are essential for domains like IoT and automotive. Simply imagine being unable to update your car’s software after a critical vulnerability has been disclosed, or of failing to migrate your data to your new car!
The core challenge lies in how the enclave state is sealed: it is typically encrypted with a key tied to a specific processor and enclave binary. This tight binding makes secure state transfer inherently difficult. As of today, the only way to perform updates or migrations is to extract the code, state, and keys from the TEE in plaintext, manually apply updates or carry out migration, and then re-provision the TEE with the modified state. This process opens multiple attack vectors---malware could exfiltrate keys, leak state, or tamper with the software---ultimately undermining the very security guarantees a TEE is supposed to provide.

In this work, we introduce \lib, a new toolkit that extends RISC-V TEEs with comprehensive lifecycle management capabilities. Built atop the popular Keystone---an open-source framework for customized RISC-V TEEs---yet designed to remain compatible with other RISC-V TEE frameworks such as CURE \cite{DBLP:conf/uss/BahmaniBDJKSS21/cure} and ACE \cite{hasp/Ozga23/ace}, \lib bridges the gap between static enclave execution and the dynamic operational needs of modern embedded systems.
Specifically, \lib extends the Security Monitor (SM)---the trusted firmware layer of RISC-V TEEs---with features that are particularly essential for the software development lifecycle and complements them with a user library that exposes a simple and intuitive API for application developers, ensuring seamless integration into enclave applications. We instantiate \lib with three essential modules providing support for secure enclave-local timekeeping, state continuity, secure software updates and migrations, allowing enclaves to evolve securely across software versions and hardware devices.
The modular design of our toolkit enables firmware developers to selectively integrate these components, minimizing the TCB and runtime overhead. 

We analyze the security of \lib and show that it enables secure, low-latency update and migration operations that allow secure state transfer across enclave versions and devices while preserving key security properties such as atomicity, authenticity, and state continuity.

Our implementation adds only a modest increase of approximately 750 LoC to the trusted computing base (TCB): 50 for trusted time, 160 for state continuity, and 540 for update and migration support. The runtime memory overhead is similarly low: the update and migration module requires just 4 B of additional memory per enclave and 232 B per active update or migration.
Moreover, our evaluation results show that our approach introduces negligible runtime overhead while maintaining practical performance. Service downtime is limited to 10ms for updates and 0.7s for migrations of enclaves with up to 16KB of state. 
This suits well deployment scenarios where brief service suspension is acceptable, such as parked vehicles or idle IoT devices, and is consistent with current automotive practices, where over-the-air (OTA) updates are routinely delivered via mobile networks or Wi-Fi and typically require several minutes to complete~\cite{adesso2025ota}.

By enabling secure and flexible lifecycle operations, \lib fills a critical gap in the RISC-V TEE ecosystem. 
\lib is publicly available at~\cite{our_artifact}, and we hope that its open and extensible structure encourages community adoption and further extensions, paving the way for broad deployment of RISC-V TEEs in automotive and IoT ecosystems.

\section{Overview of RISC-V-based TEEs}
\label{sec:background}

Trusted Execution Environments (TEEs) leverage hardware mechanisms to restrict memory access at runtime, providing isolated environments for executing sensitive user code. These environments typically support either process-based isolation, where individual user processes are protected in enclaves~\cite{DBLP:conf/isca/HoekstraLPPC13/sgx,whitepaper:trustzone,eurosys/LeeKSAS20/keystone,uss/CostanLD16/sanctum}, or virtual machine-level isolation, where trusted virtual machines (TVMs) are isolated from the rest of the system~\cite{whitepaper:tdx,whitepaper:sevSnp,hasp/Ozga23/ace}. While several commercial TEEs exist, most are proprietary and closed source.

In contrast, the RISC-V open-source instruction set architecture (ISA) has enabled a diverse landscape of customizable TEEs~\cite{uss/CostanLD16/sanctum,eurosys/LeeKSAS20/keystone,hasp/Ozga23/ace,penglai_doc,DBLP:conf/uss/BahmaniBDJKSS21/cure, DBLP:conf/aspdac/PanPMZZY0LX025/Dep-TEE}. 
The ISA defines three core privilege levels: user mode (U-mode) for executing regular application code, supervisor mode (S-mode) for operating system services, and machine mode (M-mode) for privileged firmware and hardware control. Optionally, the hypervisor extension introduces an additional hypervisor mode (H-mode), positioned between M-mode and S-mode, to support virtualization. Physical Memory Protection (PMP) is a key RISC-V feature, enabling M-mode software to configure memory access policies and enforce isolation.
In what follows, we provide an overview of prominent RISC-V-based TEE architectures.

\subsection{RISC-V TEEs}

\begin{figure}
    \centering
    \includegraphics[width=1.0\linewidth]{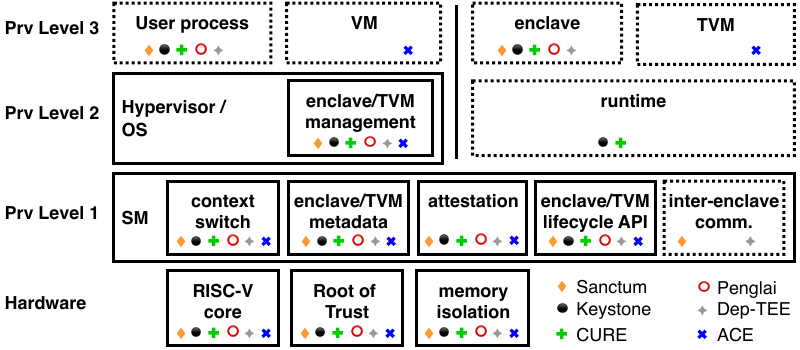}
    \caption{Overview of the system components of various RISC-V-based TEE architectures. The symbols mark the architecture in which each component is present. Components that are not common to all architectures are dashed.}
    \label{fig:common_denominator}
\end{figure}

RISC-V TEEs use a software component known as the Security Monitor (SM), operating in M-mode. The SM manages enclave lifecycle events, enforces memory protection, handles secure context switching, and provides attestation by measuring the initial enclave state and signing it with a device-specific key, allowing to verify that code was deployed unmodified in a genuine TEE. We present the specifics of RISC-V-based TEEs below. \Cref{fig:common_denominator} provides an overview of the components featured in these RISC-V TEEs.

\vspace{0.2cm}
\subheading{Sanctum.} Sanctum~\cite{uss/CostanLD16/sanctum} is a process-level TEE that isolates user-mode applications. It assumes a hardware root of trust that includes boot firmware, a measurement root, and a protected attestation key. Memory isolation is enforced via customized page coloring and translation lookaside buffer (TLB) management. The M-mode Security Monitor mediates access to enclave resources and enables secure inter-enclave communication using shared memory mailboxes. Although Sanctum introduces a hypervisor running in H-mode, it is not implemented in the prototype. The SM exposes an interface to the OS for enclave initialization, execution, and destruction. Enclaves manage their own page tables and rely on a provided runtime and modified standard library for abstraction. Attestation is delegated to a dedicated enclave, which receives the attestation key from the SM.

\vspace{0.2cm}
\subheading{Keystone.} Keystone~\cite{eurosys/LeeKSAS20/keystone} extends this design with a modular architecture, allowing to tailor the TEE to specific application requirements. Instead of relying on custom memory isolation hardware, Keystone uses RISC-V's PMP extension to isolate physical memory regions. Each enclave consists of a user-mode application and a supervisor-mode runtime. The runtime handles virtual memory management, multithreading, and host interaction through shared memory. This structure allows legacy applications to run inside enclaves with minimal changes. Keystone also includes a sealing mechanism that derives encryption keys from the enclave and device identity for secure persistent storage in untrusted memory.

\vspace{0.2cm}
\subheading{CURE.} CURE~\cite{DBLP:conf/uss/BahmaniBDJKSS21/cure}  
supports multiple enclave types, including user-mode and kernel-mode enclaves, including the Security Monitor itself. It optionally supports a hypervisor in H-mode, although this is not implemented in the prototype. One of CURE's distinguishing features is its use of globally unique enclave labels that persist across software versions, allowing to prevent re-installation of outdated software. It introduces hardware extensions to support context isolation, enforce bus-level access control, and manage cache partitions dynamically. Like Sanctum, CURE includes enclave page tables within the enclave’s memory and relies on the SM for lifecycle management and attestation. Metadata structures maintained by the SM include sealing keys and version identifiers.

\vspace{0.2cm}
\subheading{PENGLAI.}  PENGLAI~\cite{penglai_doc}  
is optimized for serverless environments with short-lived and stateless enclaves. It supports fast creation by forking from \textit{shadow enclaves} for scalability and high performance. Memory isolation is enforced using \textit{Guarded Page Tables} and a \textit{Mountable Merkle Tree} with cache-line locking and RISC-V PMP.

\vspace{0.2cm}
\subheading{Dep-TEE.} Dep-TEE~\cite{DBLP:conf/aspdac/PanPMZZY0LX025/Dep-TEE}  
extends Penglai with efficient inter-enclave communication. It enforces isolation and integrity via S-mode PMP (SPMP), cache line locking, and Merkle trees. The SM manages enclave metadata and page tables, providing APIs for enclave creation, termination, suspension, resumption, and attestation. While Penglai is limited to short-lived functions, Dep-TEE extends its capabilities to more complex, stateful workloads.

\subheading{ACE.} ACE~\cite{hasp/Ozga23/ace,arxiv/OzgaHLGSPJD25/ace_latest}  
diverges from other RISC-V TEEs by targeting confidential virtual machines instead of processes. It leverages the RISC-V hypervisor extension and runs the hypervisor in H-mode, delegating S- and U-mode to guest VMs. The M-mode SM handles secure context switching, Trusted VM (TVM) metadata, and attestation. Memory isolation is achieved through PMP and two-level address translation. Untrusted hypervisors manage all VMs, invoking SM APIs to promote them to TVMs, and to run or destroy them. ACE emphasizes formal verification of its SM and offers strong correctness and isolation guarantees for VM-based workloads.

\section{System Model \& Requirements}
\label{sec:system_model_requirements}

In this section, we describe the system model and the design requirements for our toolkit.

\subsection{System \& Threat Model}
\label{sec:system_model}

\subheading{Motivation.} 
We consider an automotive setting, where several ECUs embed a TEE~\cite{blog:embitel/automotive/TEEvsHSM,blog:embitel/automotive/TEEinECU}. Modern vehicles increasingly rely on digital components for safety, functionality, and business logic, such as automated payment systems for tolls and parking, features gated behind subscription models, or performance restrictions such as software-enforced speed limits. Here, TEEs play a central role in preventing tampering, for instance, by stopping drivers from bypassing paywalls or modifying the speedometer, without resulting in considerable computational overhead. Some of this functionality requires access to a trusted source of time. For example, features billed by usage duration require a local timer that faithfully reports the enclave’s runtime; otherwise, a user could underreport usage.

Automotive software additionally faces lifecycle and data continuity challenges. ECUs may reach end-of-life before the enclave applications they host, requiring secure migration of enclave state (such as active subscriptions or user credentials) to replacement units. Car-sharing scenarios---whether in corporate fleets or public mobility services---similarly demand transferring personal data between vehicles, while rapid and secure rollout of updates is essential---both to patch vulnerabilities and to introduce new features. 
Although RISC-V is not yet widely deployed in production vehicles, its adoption is only expected to increase in the near future. Namely, Mobileye plans to roll out RISC-V-based EyeQ Ultra silicon between 2027 and 2029, targeting a majority of RISC-V-based ADAS systems by 2030~\cite{riscvForAutomotiveAI}, and Infineon has announced an automotive RISC-V MCU family in 2025~\cite{infineonRiscvForAutomotive}, with reference architectures developed in collaboration with Bosch, Nordic Semiconductor, NXP, and Qualcomm~\cite{quintauris}. While current ECUs already integrate TEEs for safety-critical functions~\cite{blog:embitel/automotive/TEEvsHSM,blog:embitel/automotive/TEEinECU,autocryptTee}, RISC-V platforms currently lack comprehensive support for trusted timers, secure state migration, and enclave updates---essential building blocks required for a full-stack RISC-V automotive solution.

\begin{figure}[tbp]
    \centering
    \includegraphics[width=1.0\linewidth]{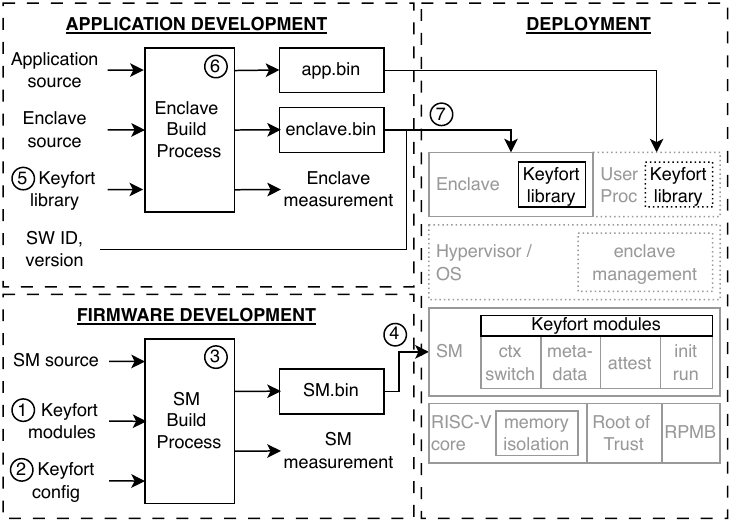}
    \vspace{-0.2cm}
    \caption{System model capturing the interaction of \lib with the ecosystem. The \lib modules \protect\circled{1}-\protect\circled{2} are integrated into the SM binary during build \protect\circled{3}-\protect\circled{4}. Enclave applications then include the \lib library \protect\circled{5}-\protect\circled{7}. Untrusted system components are dashed.}
    \label{fig:system_model_env}
\end{figure}

\vspace{0.2cm}
\subheading{System.} To remedy this, we aim in this work \emph{to design and implement a toolkit for RISC-V TEEs} that encompasses the following components:
a trusted time module, a state continuity module, and a secure update and migration module. Based on the discussion in \Cref{sec:background}, we reckon that all RISC-V-based TEEs share a common set of components---such as a root of trust, a RISC-V core with memory isolation support, and a Security Monitor---as summarized in \Cref{fig:system_model_env}. Given the heavy reliance on these components, we believe that future designs catered to the automotive industry are also likely to follow similar design patterns. Furthermore, many vehicles have access to a replay-protected memory block (RPMB), which is present in standard eMMC and UFS devices~\cite{article:armSecureAutomotiveRPMB}. 

We adopt a modular design, implementing each feature provided by \lib as a separate module (i.e., source file) within the SM (Step~\circled{1}). 
The modules extend existing SM functions and data structures, and define new functions within the SM. They are accompanied by a configuration file (Step~\circled{2}) for configuring which modules should be included and some module-specific parameters. This approach allows us to include only the required modules, keeping the TCB minimal and complying with the modular design of Keystone~\cite{eurosys/LeeKSAS20/keystone} and Rust-based SMs~\cite{hasp/Ozga23/ace}. Firmware providers can simply include the modules and the config file in the source directory of any RISC-V TEE's SM together with the vendor's public key certificate, requiring only one include directive to include the config file in the SM's enclave file. When compiling the SM's source code (Step~\circled{3}), the \lib modules are directly embedded into the SM's binary, exposing a set of SBI calls to lower privileged software as an API for efficient usage of \lib's functionality when deployed on the RISC-V core (Step~\circled{4}). We opt to incorporate \lib's core functionality into the SM, as some of its functionality---such as accessing timer registers---requires SM privileges, necessitating an extension or modification of the SM. Furthermore, the SM can provide this functionality with lower latency if the function calls are directly trapped in the SM.

The last component of \lib is a user library (Step~\circled{5}), that is included in the enclave and application code during compilation (Step~\circled{6}) and serves as a wrapper for convenient calling of \lib's API. Specifically, this library exposes functions to query the enclave runtime, to update enclave code, and migrate enclave state between two enclaves of different versions or host devices. The resulting enclave binary is then tied to a software ID and version number via a certificate and deployed on the target device (Step~\circled{7}).

During updates and migrations, the communication between the SMs of target ECUs and the car manufacturer or owner is protected by a secure channel with encryption and a MAC for all exchanged messages (e.g., TLS), following standard practices for secure communication. All participants hold asymmetric key pairs $(sk,pk)$ and mutually authenticate using known public keys.

\subheading{Threat model.} Consistent with prior work~\cite{eurosys/LeeKSAS20/keystone,DBLP:conf/uss/BahmaniBDJKSS21/cure,uss/CostanLD16/sanctum,penglai_doc,DBLP:conf/aspdac/PanPMZZY0LX025/Dep-TEE,hasp/Ozga23/ace}, we assume that the SM and the device's hardware components are part of the Trusted Computing Base (TCB), and cannot be compromised by the adversary. In this paper, we focus on attacks originating from untrusted privileged software (e.g., an OS or hypervisor provided by a manufacturer) or from a local attacker physically controlling the device (e.g., the owner of an IoT device or vehicle).

\subsection{Design Requirements}
\label{sec:requirements}

To the best of our knowledge, there exists no toolkit that supports the end-to-end life cycle management in the maintenance phase of enclave applications with the functionalities described above.
Given the sensitive nature of enclaves, these functional capabilities must be accompanied by strong security guarantees. We define the following security requirements to ensure the security of the update and migration processes. 

\begin{description}[leftmargin=0.2cm]
    \item[S1---Authenticity.] Only trusted entities may initiate updates or migrations. The source may migrate only to trusted destinations, and destinations may accept enclaves or states only from trusted sources.
    
    \item[S2---Code Integrity.] Any unauthorized modification of the enclave code during update or migration must be detectable.
    
    \item[S3---State Confidentiality and Integrity.] The enclave state must remain confidential during update or migration, and any unauthorized modification must be detectable.
    
    \item[S4---Software Rollback Protection.] After installing version~$v$ of the enclave software, any prior version~$v' < v$ must be considered invalid and rejected to prevent re-installation of potentially vulnerable code.
    
    \item[S5---Atomicity.] Updates and migrations must be atomic: either the process completes successfully with only the destination enclave active, or it fails with only the source enclave active. 

    \item[S6---State Continuity.]  The continuity of the enclave's state must be maintained at all times, ensuring it remains consistent and uncorrupted throughout regular enclave operation and updates or migrations. 
\end{description}

\noindent
In addition to functional and security guarantees, the system must also operate efficiently, especially in resource-constrained environments such as edge devices or embedded systems. The following performance requirements aim to ensure that update and migration procedures do not impose prohibitive costs.

\begin{description}[leftmargin=0.2cm]
    \item[P1---Overhead.] The communication, computational, and memory overheads of the update and migration protocols must be minimal to support resource-constrained platforms.
    
    \item[P2---Minimal Downtime.] The enclave service must experience minimal interruption during updates or migrations to maintain high availability.
\end{description}

\section{\lib Components \& Specification}
\label{sec:components}

In \Cref{sec:requirements}, we defined three crucial features for secure enclave lifecycle management, namely a trusted source of time, state continuity support, secure software updates, and secure software migrations. In this section, we present a toolkit, dubbed \lib, to address these requirements. We start by describing the set of extensions for the Security Monitor that \lib comprises. We then show how to leverage \lib for secure enclave migrations and updates by proposing new migration and update protocols based on \lib's API.

\subsection{Security Monitor Extensions}
\label{sec:sm_extensions}

The Security Monitor is the M-mode software that enforces TEE guarantees. To do so, it exposes APIs for enclave initialization, execution, and destruction to the untrusted OS or hypervisor. During initialization of an enclave \E with the binary \bin, the SM initializes a data structure to track enclave metadata---typically a \emph{struct}---including a local enclave identifier \eid and the enclave's measurement \meas, among other architecture-specific information relevant for secure enclave management. Moreover, it provides enclaves with an attestation interface returning a signed certificate of \meas to prove that the enclave runs the correct binary inside a genuine TEE.

In what follows, we present \lib, which comprises a set of modifications and extensions to this basic SM functionality that are required to provide secure update and migration, state continuity, and trusted time features for RISC-V TEEs.
Designing these extensions is not straightforward, as they significantly increase the complexity of the Security Monitor’s trusted code base. Namely, the SM must remain minimal while coordinating enclave updates and migrations under an untrusted OS. This entails securely managing enclave lifecycle transitions, maintaining rollback-resistant versioning state, and ensuring atomic state continuity across devices---all within M-mode, without relying on external trust. 

We start by introducing five modifications (Mod. 1-Mod. 5) to existing data structures and functions:

\begin{description}[leftmargin=0.2cm]
    \item[Mod. 1:] The \texttt{enclaves} data structure tracks all enclave's $eid$s and measurements \meas. We modify it to additionally track the following information per enclave:
    \begin{itemize}
        \item \ID: Software identifier (common across versions).
        \item \ver: Version number of the binary \bin.
        \item $N$: Number of allowed concurrent instances with the same measurement.
        \item \enclTime: Enclave runtime until the last enclave exit.
        \item \entryTime: Timestamp of the last enclave entry.
        \item \resumeok: A boolean flag specifying if the enclave can be started/resumed.
    \end{itemize}
    
    \item[Mod. 2:] The \texttt{init} function initializes a new enclave. It takes a binary \bin as an input, and initializes a new enclave with it. The function assigns a unique \eid to the enclave and initializes an enclave metadata entry accordingly. Finally, it returns \eid to the caller. 
    We extend the function's signature to additionally receive \ID, \ver, and $N$ as inputs. Our modification performs two additional checks before initializing the enclave: (1) it verifies that the enclave software is up to date by checking that $v$ matches the last installed version (\vlate) for the enclave's \ID stored in Ext. 1---a new entry is created if it does not yet exist---and (2) the number of initialized enclaves $n$ with the same \meas satisfies $n<N$. The default value for $N$ is 1. The step (1) of the verification can be bypassed if the enclave's \ID is contained in a list of scheduled updates and migrations (cf. Ext. 3). In this case, the entry is removed from the list of scheduled updates/migrations and the \resumeok flag is set to \emph{false}. If all checks pass, the function initializes the enclave, and the additional information is included in the enclave metadata entry accordingly. 

    \item[Mod. 3:] The \texttt{run/resume} functions takes an \eid as an input and starts or resumes the execution of the enclave with the corresponding \eid at a given entry point. We extend this function(s) with a check to validate the \resumeok flag, denying the execution of enclaves where this flag is set to \emph{false}.

    \item[Mod. 4:] The \texttt{context\_switch} function securely switches the execution context between the trusted enclave and the untrusted host application, clearing the processor state where necessary. We incorporate an additional step in this function(s), probing the \mtime register during context switches, yielding a timestamp $t$. 
    If the switch is from untrusted software to the enclave, the SM sets the enclave entry time $\entryTime=t$ in the enclave metadata; otherwise, it updates the enclave's execution time: $\enclTime=\enclTime+(t-\entryTime)$. Additionally, we extend this function to again validate the \resumeok flag before switching execution to the enclave.

    \item[Mod. 5:] The \texttt{attest} function creates an attestation report, signing the enclave measurement and selected metadata with a secret device key. We extend it to include three new fields from the enclave metadata in the attestation report: \ID, \ver, and $N$.
\end{description}

\noindent
Beyond the described modifications of existing data structures and functions, \lib encompasses 15 extensions (Ext.~1-Ext.~15), introducing four new data structures and eleven additional functions to the SM's API, to support secure software lifecycle development (cf. \Cref{sec:system_model_requirements}): 

\begin{description}[leftmargin=0.2 cm]
    \item[Ext. 1:] The \texttt{sw\_versions} list contains a tuple $(\ID, \vlate)$ for each enclave software that was installed on the device at some point in time, tracking the latest installed software version and the corresponding enclave measurement. While the enclave metadata is volatile and only kept in runtime memory, this versioning information must be persisted in the SM's rollback-resistant non-volatile memory (i.e., RPMB) to prevent software rollback attacks after device reboots.

    \item[Ext. 2:] The \texttt{monotonic\_counters} list stores monotonic counters for enclaves. Each monotonic counter is defined by a triple of the counter ID $ctr_{ID}$, its value $ctr_{val}$, and the software \ID of the associated enclave.

    \item[Ext. 3:] The \texttt{scheduled\_migrations} list stores enclave $ID$s that are scheduled to be updated or migrated, allowing to bypass version checks during initialization (cf. Mod. 2).

    \item[Ext. 4:] The \texttt{migrations} list stores metadata for all active update and migration processes, each entry tracking relevant information for a state migration from a source enclave \ESrc running on SM \SMSrc to a destination enclave \EDst on \SMDst:

    \begin{itemize}
        \item \ID: Software identifier for \ESrc and \EDst.
        \item $target\_op$: \emph{update}, or \emph{migration source/destination}.
        \item \eidSrc: Local enclave identifier of \ESrc.
        \item \eidDst: Local enclave identifier of \EDst.
        \item \measSrc: Measurement of \ESrc.
        \item \measDst: Measurement of \EDst.
        \item \pkSrc: Public key of \SMSrc.
        \item \pkDst: Public key of \SMDst.
        \item \seed: Seed for deriving the symmetric migration key.
        \item $T_{SM}$: Timeout for the state migration.
    \end{itemize}
    
    \item[Ext. 5:] The \texttt{enclave\_local\_time} function returns the number of \mtime ticks that the CPU spent in the calling enclave's execution context. Therefore, it probes the \mtime register yielding $t$. This register is only accessible to M-mode software, ensuring that the OS and other untrusted software cannot tamper with the timestamp. The function then computes the enclave runtime as $\enclTime=\enclTime+(t-\entryTime)$ and returns it to the caller. The platform must provide a mechanism to determine the period of an \mtime tick~\cite{riscVmanual}, allowing it to translate the register's value into a concrete timestamp if necessary. 
    
    \item[Ext. 6:] The \texttt{schedule\_migration} function schedules an enclave for migration. 
    Recall that the enclave's lifecycle is managed by the untrusted OS or hypervisor (cf. \Cref{fig:common_denominator}). Thus, the SM cannot initialize a new enclave by itself. However, the destination enclave of a migration must be disabled at initialization to satisfy security requirements S5 (atomicity) and S6 (state continuity)---the OS cannot be trusted to start the destination enclave only after the source has been stopped---diverging from standard enclave initialization. At the same time, exposing a new lifecycle function for initializing a migration's destination enclave to the operating system leaks information about the migration process. 
    Therefore, this function performs necessary checks and schedules an enclave \ID for migration, such that the SM can disable execution when an enclave with this \ID is subsequently initialized (cf Mod. 2).
    It receives the software \ID as an input. The function verifies that \ID is not scheduled for update or migration and that no enclave with \ID exists on the device, subsequently adding \ID to the list of scheduled updates/migrations. As \ID is only removed from the list of scheduled updates/migrations upon initialization, a concurrent migration request for the same \ID is rejected.
    
    \item[Ext. 7:] The \texttt{schedule\_update} function schedules an enclave for update, representing the update counterpart to Ext. 6. It receives the enclave's software \ID and its version \ver as inputs. The function verifies that \ID is not scheduled for update or migration and that exactly one enclave with \ID and a version $<\ver$ exists on the device, subsequently adding \ID to the list of scheduled updates/migrations.
    
    \item[Ext. 8:] The \texttt{state\_migration} function initiates state migration between two enclaves, potentially residing on distinct devices (i.e., SMs). It receives as inputs two public keys of the two SMs (\pkSrc and \pkDst), the enclaves' identifiers (\eidSrc and \eidDst), and their measurements (\measSrc and \measDst). This function validates that one of the included public keys matches its own and uses its position to determine whether it is the source or destination of the state migration. Subsequently, the SM verifies that (i) the enclave $E_X$ identified by $eid_X$ has the correct measurement $m_X$ and (ii) no other state migration is scheduled for $eid_X$. If $\pkDst\neq\pkSrc$ (indicating a migration), the SM additionally verifies that $\measSrc=\measDst$. Otherwise, in the case of an update, it confirms that the enclaves' software identifiers match (i.e., $\IDSrc=\IDDst$) and that the destination enclave is of a higher software version: $\vSrc<\vDst$. Finally, the SM initializes a migration metadata entry.
    Here, \seed is left empty and  $T_{SM}$ is initialized with a timeout.
    
    \item[Ext. 9:] The \texttt{get\_transport\_key} function generates a transport key for the migration of the calling enclave's state. It validates that a migration for the calling enclave is registered. It then derives the transport key from the involved enclave measurements and a seed \seed: $\key=H(\seed,\measSrc,\measDst)$. If $target\_op$ indicates an update, \seed is sampled at random. Otherwise, \seed equals the session key for the communication between \SMSrc and \SMDst. This key derivation enforces several security guarantees at once before providing \EDst access to the state: (1) it binds \key to a specific session, ensuring state continuity (S6) in case an update/migration failed and is restarted while \ESrc may have updated it state in the meantime; (2) it binds \key to the enclave identities of \ESrc and \EDst, enforcing code integrity; and (3) it enforces state confidentiality, allowing only a specific set of enclaves on a specific set of devices to access the state in a specific session.
    Finally, the SM returns \key to the enclave.
    
    \item[Ext. 10:] The \texttt{execution\_switch} function switches the execution from the enclave identified by \eidSrc to the one identified by \eidDst. Specifically, it validates that migrations entry exists for the two enclaves. It then sets the \resumeok flag of \eidSrc to \textit{false} (if $target\_op$ indicates an update or migration source) and sets the one of \eidDst to \textit{true} (if $target\_op$ indicates an update or migration destination). 
    This function is crucial to satisfy security requirement S5 (atomicity), as it ensures that \ESrc and \EDst cannot run in parallel. Note that one could also rely on the untrusted host to first stop \ESrc and then resume \EDst. However, this choice would increase communication overhead and impose tracking overhead on the SM, ensuring that \ESrc has been stopped before \EDst is started to provide the same security guarantees.
    
    \item[Ext. 11:] The \texttt{migration\_commit} function finalizes a state migration and clears metadata. It verifies that the calling enclave's \eid matches \eidDst of the corresponding metadata entry. In case of an update, the SM first updates the version associated with \ID, setting it to \vDst. It then destroys the corresponding source enclave (identified by \eidSrc) and clears the migration metadata. Finally, it signals successful completion of the commit to the calling enclave.
    Here, it is important to clear the migration metadata only \emph{after} the source enclave has been destroyed to guarantee atomicity (S5), even in the presence of crashes. 

    \item[Ext. 12:] The \texttt{allocate\_MC} function allocates a new monotonic counter. It initializes the \ID field of the monotonic counter to the calling enclave's \ID and returns the corresponding $ctr_{ID}$. Binding the monotonic counter to the software ID of an enclave provides state continuity (S6) guarantees across software versions.
    
    \item[Ext. 13:] The \texttt{get\_MC\_value} function queries the value of a monotonic counter, receiving the counter ID $ctr_{ID}$. It verifies that the calling enclave's \ID matches the counter's \ID field and returns the corresponding counter's $ctr_{val}$.  
    
    \item[Ext. 14:] The \texttt{inc\_MC} function increments the value of a monotonic counter, receiving the counter ID $ctr_{ID}$. It verifies that the calling enclave's \ID matches the counter's \ID field, increments and returns the corresponding counter's $ctr_{val}$.  
    
    \item[Ext. 15:] The \texttt{free\_MC} function frees a monotonic counter, receiving the counter ID $ctr_{ID}$. It verifies that the calling enclave's \ID matches the counter's \ID field and removes the monotonic counter from the monotonic counter list.
   
\end{description}

In the following sections, we describe how we leverage these extensions to satisfy the requirements specified in \Cref{sec:requirements}. \Cref{tab:extensions} provides an overview of the modifications and extensions used by each individual module.

\newcommand{\extRow}[5]{ \rule{0.5em}{0pt} \textbf{#2} & #1 & #3 & #4 & #5 \\ }

\begin{table}[tbp]
    \centering
    \footnotesize
    \caption{Overview of the SM modifications and extensions in our toolkit, for each one reporting which modules use it.}

    \scalebox{0.9}{
    \begin{tabular}{llccc}
        \toprule
         & & \multicolumn{3}{c}{\textbf{Module}} \\ \cmidrule{3-5}
         & & \textbf{Trusted} & \textbf{State} & \textbf{Update \&}  \\
         \textbf{Component} & \textbf{Description} & \textbf{Time} & \textbf{Continuity} & \textbf{Migration}  \\ 
         \midrule

        \multicolumn{5}{l}{\textbf{Modifications}} \\
        \extRow{Enclave metadata}{Mod. 1}{\cmark}{\cmark}{\cmark}
        \extRow{Initialize enclave}{Mod. 2}{\xmark}{\cmark}{\cmark}
        \extRow{Run/resume enclave}{Mod. 3}{\xmark}{\xmark}{\cmark}
        \extRow{Context switch}{Mod. 4}{\cmark}{\xmark}{\xmark}
        \extRow{Attest enclave}{Mod. 5}{\xmark}{\cmark}{\cmark}
        \midrule
        
        \multicolumn{5}{l}{\textbf{Extensions}} \\
        \extRow{SW version tracking}{Ext. 1}{\xmark}{\cmark}{\cmark}
        \extRow{Monotonic counters}{Ext. 2}{\xmark}{\cmark}{\xmark}
        \extRow{SW $ID$s for update/migration}{Ext. 3}{\xmark}{\xmark}{\cmark}
        \extRow{Migration metadata}{Ext. 4}{\xmark}{\xmark}{\cmark}
        \extRow{Get enclave local timestamp}{Ext. 5}{\cmark}{\xmark}{\xmark}
        \extRow{Schedule \ID for migration}{Ext. 6}{\xmark}{\xmark}{\cmark}
        \extRow{Schedule \ID for update}{Ext. 7}{\xmark}{\xmark}{\cmark}
        \extRow{Initialize state migration}{Ext. 8}{\xmark}{\xmark}{\cmark}
        \extRow{Get migration transport key}{Ext. 9}{\xmark}{\xmark}{\cmark}
        \extRow{Switch execution (\ESrc to \EDst)}{Ext. 10}{\xmark}{\xmark}{\cmark}
        \extRow{Finalize migration}{Ext. 11}{\xmark}{\xmark}{\cmark}
        \extRow{Allocate monotonic counter}{Ext. 12}{\xmark}{\cmark}{\xmark}
        \extRow{Get monotonic counter value}{Ext. 13}{\xmark}{\cmark}{\xmark}
        \extRow{Increment monotonic counter}{Ext. 14}{\xmark}{\cmark}{\xmark}
        \extRow{Free monotonic counter}{Ext. 15}{\xmark}{\cmark}{\xmark}
        \bottomrule
    \end{tabular}}
    \label{tab:extensions}
\end{table}

\subsection{Trusted Time}
\label{sec:time}

We start by describing the trusted time source function of our toolkit. This component enables enclaves to query a trusted global timestamp and a local enclave timestamp. It leverages the \mtime register of the RISC-V ISA. This register is a memory-mapped register which is accessible only to M-mode software and increments at a constant frequency (e.g., once per clock cycle)~\cite{riscVmanual}. The period of an \mtime tick is platform dependent, but constant, allowing conversion into a concrete timestamp. 

Note that the \mtime register is a read-write register that is accessible to \emph{all} M-mode software. Hence, not only the trusted SM but any firmware running in M-mode can set \mtime to an arbitrary value, violating its monotonicity. While CURE~\cite{DBLP:conf/uss/BahmaniBDJKSS21/cure} is the only RISC-V TEE considering threats from concurrently running firmware, such attacks can be mitigated by allowing only the SM to access the \mtime register. For instance, Kuhne \etal~\cite{DBLP:conf/uss/KuhneVS25/Dorami} propose an approach to achieve intra-M-mode privilege separation. In particular, they assign each M-mode firmware an individual PMP region and prohibit any non-SM code from invoking PMP instructions, thereby preventing modifications of the PMP registers. Hence, we adopt a similar approach where the SM creates a PMP entry during boot that covers the \mtime register's memory address and assigns it to its own memory. As a result, only the SM is enabled to tamper with the \mtime value. Since the SM is part of the TCB, probing this register yields a trusted timestamp $t$.

The \texttt{time} register is a read-only shadow of \mtime, and can be read by lower privileged software using the \texttt{rdtime} instruction. Hence, enclaves can use the \texttt{rdtime} instruction to retrieve a trusted global timestamp, representing the number of \mtime ticks since the last device restart. This value can be converted to a human-interpretable form by accounting for the period of an \mtime tick.

\begin{figure}[tbp]
    \centering
    \includegraphics[width=0.92\linewidth]{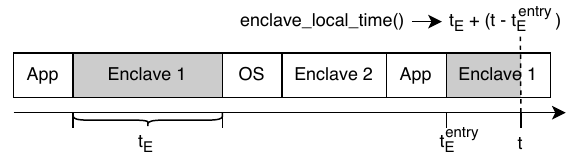}
    \caption{Overview of enclave-local time operation. The SM tracks \enclTime and \entryTime during context switches (App$\rightarrow$Enclave 1 and Enclave 1$\rightarrow$OS). Grey areas indicate Enclave 1’s active runtime at time $t$ when it queries its local time.}
    \label{fig:trusted_time}
\end{figure}

In contrast, tracking an enclave's runtime relies on a subset of the SM modifications described in \Cref{sec:sm_extensions}. Specifically, the trusted time source component of our toolkit includes the function \texttt{enclave\_local\_time} (Ext. 5) and the associated \entryTime and \enclTime fields of the enclave metadata (Mod. 1), as well as the \linebreak \texttt{context\_switch} extensions (Mod. 4). \texttt{Enclave\_local\_time} is exposed via an SBI call to the enclave, enabling it to determine the number of \mtime ticks spent in the enclave's execution context since it was initialized on the corresponding platform (cf. \Cref{fig:trusted_time}).  

\subsection{State Continuity}
\label{sec:rollback_cloning}

\begin{figure}[tbp]
    \centering
    \includegraphics[width=0.8\linewidth]{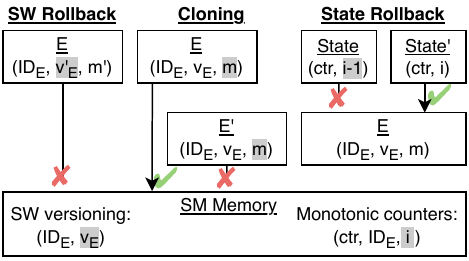}
    \caption{Overview of the state continuity module. Cross marks indicate enclave initializations or state imports aborted due to mismatches with the SM (shown in gray), while check marks denote successful operations.}
    \label{fig:state_continuity}
\end{figure}

In the following, we describe the state continuity mechanisms of \lib, which collectively mitigate software rollback, cloning, and rollback-based forking attacks on the enclave state. These mechanisms (cf.~\Cref{fig:state_continuity}) are required to satisfy the security requirements S4 (software rollback protection) and S6 (state continuity) beyond enclave updates and migrations, as presented in Sections~\ref{sec:update} and~\ref{sec:migration}.

\vspace{0.2cm}
\subheading{Software rollback protection.}
In a software rollback attack, an adversary re-installs an outdated enclave binary---potentially containing known vulnerabilities---to regain access to data sealed under that version.
The practical impact of such attacks was demonstrated on Arm TrustZone by Busch \etal~\cite{DBLP:conf/uss/BuschMP24/spillTheTeA}.

To prevent this, we require the SM to maintain a persistent, rollback-protected record of the latest version observed for each software identifier. Specifically, the SM stores tuples $(\ID, \vlate)$ in encrypted, non-volatile memory (cf. Ext. 1). During enclave initialization, the SM checks whether an entry for the given \ID exists and validates the version of the enclave to be initialized against \vlate  (cf. Mod. 2). 
If $\ver<\vlate$, the SM detects a software rollback attempt. If $\ver>\vlate$, it identifies an unauthorized upgrade attempt, which must instead follow the update protocol to maintain state continuity. In both cases, the initialization request is rejected. This mechanism ensures that only a precisely defined version of an enclave software can be executed.

To improve transparency, we extend the attestation report to include both \ID and \ver. Moreover, $\vlate$ is updated to \vDst during Steps~\circled{4m} and \circled{4n} of the update and migration protocols (cf. Ext.~11).

\vspace{0.2cm}
\subheading{Cloning prevention.}
Version tracking prevents the simultaneous execution of different enclave versions but does not prevent multiple concurrent instances of the same version. This limitation enables \emph{cloning attacks}, in which two identical enclaves ($E, E'$) process different inputs ($i_1, i_2$), resulting in diverging internal states ($S, S'$). Such state inconsistency undermines the assumption of continuity and can lead to security compromises~\cite{DBLP:conf/acsac/BriongosKSW23/clonebuster,DBLP:conf/ccs/NiuPZZ22/Narrator,DBLP:conf/ndss/WildeGSK25/OurNDSSForkingWay, DBLP:conf/dsn/BrandenburgerCL17/LCM}.

To prevent cloning, we require the SM to extend its enclave initialization checks to ensure that no identical enclave measurement \meas is already active. If a match is found, the SM rejects the initialization request (cf. Mod. 2).
In legitimate cases where multiple instances are desired---for example, to improve redundancy or throughput~\cite{DBLP:conf/acsac/BriongosKSW23/clonebuster}---the enclave can specify an upper bound $N$ on the number of concurrent instances. During initialization, the SM verifies that the number of active enclaves with the same \ID is strictly less than $N$; otherwise, the request is rejected.
This mechanism ensures that the enclave controls the number of concurrently active instances, preventing tampering by unauthorized clones.

\vspace{0.2cm}
\subheading{Rollback protection.}
Even in the absence of software rollbacks or cloning, \emph{state rollback attacks} can break continuity. Since enclaves cannot persist state internally, they must store encrypted data in untrusted memory using a sealing key derived from the enclave measurement and device identity. However, sealed data alone lacks freshness guarantees: an adversary can replay an older state (e.g., $S_{i-1}$ instead of the latest $S_i$), reverting the enclave to a prior state. 

Previous work on Intel SGX~\cite{DBLP:conf/uss/MateticAKDSGJC17/ROTE, DBLP:conf/ccs/NiuPZZ22/Narrator} has shown that including a monotonic counter in the sealed state provides a freshness token to detect rollbacks. While commercial TEEs rarely offer such counters, \lib extends the RISC-V SM to support software-based monotonic counters (Ext. 2).
We expose these counters via the SM API through the functions \texttt{allocate\_MC}, \texttt{get\_MC\_value}, \texttt{inc\_MC}, and \texttt{free\_MC} (\cf Ext.12-Ext. 15). Enclaves can associate a counter with their software identifier and include its value in the sealed state. Upon recovery, the enclave compares the stored counter value with the reference value obtained from the SM; any mismatch indicates a rollback attempt.
This mechanism ensures that state restoration always proceeds from the most recent enclave state.

\subsection{Secure Enclave Migration}
\label{sec:migration}

We now describe the migration protocol of our toolkit, based on the SM extensions described in \Cref{sec:sm_extensions}.
Before migration, the source enclave \ESrc runs binary \binSrc of software \IDSrc on a platform managed by the Security Monitor \SMSrc with public key \pkSrc. The enclave's current state is denoted \StateSrc. The migration is initiated by a party \party, who selects a destination platform managed by \SMDst with public key \pkDst. A new enclave \EDst is created on this destination and runs the same binary $\binDst=\binSrc$. The protocol ensures secure transfer of both binary and enclave state such that $\StateDst=\StateSrc$, while preventing concurrent execution of \ESrc and \EDst.

\begin{figure}
    \centering
    \scalebox{0.58}{

\begin{tikzpicture}[
  party/.style={draw, minimum width=2.6cm, minimum height=1cm, rounded corners},
  msg/.style={-{Latex[length=2mm]}, thick, draw=white},
  arrow/.style={-{Latex[length=2mm]}, thick},
  arrowBi/.style={{Latex[length=2mm]}-{Latex[length=2mm]}, thick},
  font=\sffamily\small,
  every node/.style={align=center},
  phase/.style={draw=black, rounded corners, inner sep=3pt, font=\bfseries\footnotesize},
  subphase/.style={draw=gray, rounded corners, inner sep=3pt, font=\bfseries\footnotesize}
]

\pgfdeclarelayer{bg}
\pgfdeclarelayer{fg} 
\pgfsetlayers{bg,main,fg}

\def\xHost{1}
\def\xSrc{4.6}
\def\xSMS{7.6}
\def\xDst{10.6}
\def\xSMD{13.6}
\def\yStep{0.9}
\def\nSteps{20.1}

\node[] (Host) at (\xHost, 0) {};
\node[party] (Src) at (\xSrc, 0) {Source\\Enclave};
\node[party] (SMSrc) at (\xSMS, 0) {Source\\SM};
\node[party] (Dst) at (\xDst, 0) {Destination\\Enclave};
\node[party] (SMDst) at (\xSMD, 0) {Destination\\SM};

\begin{pgfonlayer}{bg}
    \foreach \x in {\xSrc, \xDst, \xSMS, \xSMD} {
      \draw[dashed] (\x, -0.5) -- ++(0, -\nSteps*\yStep - 0.5);
    }
\end{pgfonlayer}

\newcommand{\msg}[4]{%
  \path (#2) ++(0,-#1*\yStep) coordinate (from);
  \path (#3) ++(0,-#1*\yStep) coordinate (to);
  \draw[msg] (from) -- (to) node[midway, above, fill=white] {#4};
  \draw[arrow] (from) -- (to) {};
}

\newcommand{\event}[3]{%
  \path (#2) ++(0,-#1*\yStep - 0.5*\yStep) coordinate (pos);
  \node[fill=white, draw=black, rounded corners, font=\scriptsize, inner sep=2pt] at (pos) {#3};
}

\msg{1.4}{Host}{SMDst}{\stepCircled{1}\texttt{schedule\_migration}(\IDDst)}
\msg{2.4}{Host}{SMDst}{\stepCircled{2}\texttt{init}(\IDDst, \vDst, \binDst)}
\msg{3.4}{SMDst}{Host}{\stepCircled{3}\eidDst}
\msg{4.4}{Host}{SMSrc}{~\stepCircled{4}\texttt{state\_migration}(\pkSrc, \pkDst, \eidSrc, \eidDst, \measSrc, \measDst)}
\msg{5.4}{Host}{SMDst}{~\stepCircled{4}\texttt{state\_migration}(\pkSrc, \pkDst, \eidSrc, \eidDst, \measSrc, \measDst)}
\event{5.4}{SMSrc}{~\stepCircled{4a}verify migration}
\event{5.4}{SMDst}{~\stepCircled{4a}verify migration}
\msg{7.0}{Host}{Src}{\stepCircled{4b}\texttt{export\_state}}
\msg{8.0}{Src}{SMSrc}{\stepCircled{4c}\texttt{get\_transport\_key}}
\msg{9.0}{SMSrc}{Src}{\stepCircled{4d}Ephemeral key $\key$}
\msg{10.0}{Src}{Host}{\stepCircled{4e}Enc state + MAC $(C, M)$}
\msg{11.0}{Host}{SMSrc}{\stepCircled{4f}\texttt{execution\_switch}(\eidSrc, \eidDst)}
\msg{12.0}{SMSrc}{SMDst}{\stepCircled{4g}\texttt{execution\_switch}(\eidSrc, \eidDst)}
\msg{13.0}{Host}{Dst}{\stepCircled{4h}\texttt{import\_state}$(C, M)$}
\msg{14.0}{Dst}{SMDst}{\stepCircled{4j}\texttt{get\_transport\_key}}
\msg{15.0}{SMDst}{Dst}{\stepCircled{4k}\key}
\msg{16.0}{Dst}{SMDst}{\stepCircled{4m}\texttt{commit\_migration}}
\msg{17.0}{SMDst}{SMSrc}{\stepCircled{4n}\texttt{commit\_migration}}
\msg{18.0}{SMSrc}{SMDst}{\stepCircled{4o}\texttt{OK}}
\msg{19.0}{SMDst}{Dst}{\stepCircled{4p}\texttt{OK}}
\msg{20.0}{SMDst}{SMSrc}{\stepCircled{4q}\texttt{OK}}
\msg{21.0}{SMSrc}{Host}{\stepCircled{5}\texttt{OK}}

\begin{pgfonlayer}{fg}
    \node[phase, fit={(0.8,-0.8) (14.8,-3.2)}] (initbox) {};
    \node[phase, fit={(0.8,-3.4) (14.8,-19.0)}] (migrationbox) {};
    \node[subphase, fit={(1,-5.8) (14.6,-13.6)}] (transferbox) {};
    \node[subphase, fit={(1,-13.8) (14.6,-18.2)}] (commitbox) {};
\end{pgfonlayer}{fg}

\node[anchor=north east, font=\bfseries\footnotesize, fill=white] at (initbox.north east) {Initialization};
\node[anchor=north east, font=\bfseries\footnotesize, fill=white] at (migrationbox.north east) {State Migration};
\node[anchor=north east, font=\bfseries\footnotesize, fill=white] at (transferbox.north east) {\color{gray}State Transfer};
\node[anchor=north west, font=\bfseries\footnotesize,fill= white] at (commitbox.north west) {\color{gray}Commit};

\end{tikzpicture} }
    \vspace{-0.3cm}
    \caption{Migration Protocol}
    \label{fig:migration_protocol}
\end{figure}
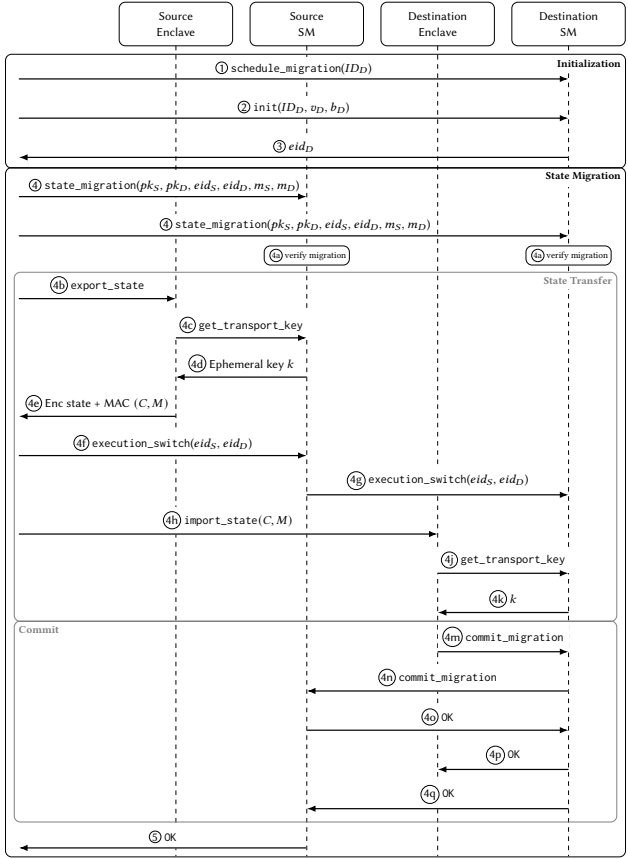

The overall migration protocol is shown in \Cref{fig:migration_protocol}, and consists of two phases: (i) initialization and (ii) state migration, which can be further decomposed into a state transfer and a commit subphase.

\subheading{Initialization phase.}
Initially, \party selects a target device to which the enclave should be migrated. It then uses the \texttt{schedule\_migration} API (Ext. 6) to send a code migration request over a secure channel to the selected destination \SM (\SMDst), specifying the software identifier \ID (\step{1}). Furthermore, \party sets a timeout $T_P$. If \party does not receive an acknowledgment from \SMSrc before $T_P$ expires, it treats the migration as failed.
\SMDst verifies that \party is authorized to initiate a migration using its public key $pk_P$ and that no enclave with the specified \ID is active. \party then requests initialization of the corresponding enclave via the normal \texttt{init} API, specifying the enclave code, including the software identifier \ID, version \ver, and binary \bin (\step{2}). The SM then creates a new enclave \EDst from \binDst. As \ID is scheduled for migration, \EDst is not yet activated and cannot process inputs. All requests to activate it are rejected. Finally, \SMDst returns \eidDst to \party (\step{3}).

\subheading{State migration phase.} 
This phase securely migrates the enclave state from \ESrc to \EDst, ensuring that neither enclave can be active at the same time. First, \party sends a state migration request to both SMs (\SMSrc and \SMDst) via the \texttt{state\_migration} API (Ext. 8), specifying their public keys and the enclave identifiers and measurements of \ESrc and \EDst (\step{4}). 
The SMs verify the migration request and set a timeout $T_{SM}$ for the migration, as specified in the function (\step{4a}).
If the migration does not complete before $T_{SM}$ expires---i.e., the migration metadata entry still exists---the SM considers the migration failed. \SMSrc resumes \ESrc, while \SMDst destroys \EDst, and both clear the metadata and notify both \party and the remote SM accordingly.
The remainder of the state migration phase is composed of the following sub-phases:

\vspace{0.5em}
\noindent\underline{State transfer phase.}
First, \party instructs \ESrc to export its state (\step{4b}). For this, \ESrc requests a fresh symmetric key \key from \SMSrc using the \texttt{get\_transport\_key} API (Ext. 9, \step{4c}). 
\SMSrc then returns \key to \ESrc (\step{4d}), which encrypts its current state \StateSrc, producing ciphertext $C$ and MAC $M$, and returns the pair $(C,M)$ to \party (\step{4e}). From this point onward, \ESrc stops processing inputs unless \SMSrc later confirms that the migration failed and \ESrc was resumed. 

Next, \party asks \SMSrc to pause \ESrc (\step{4f}) and instruct \SMDst to activate \EDst (\step{4g}), both using the \texttt{execution\_switch} API (Ext. 10). \party then forwards $(C,M)$ to \EDst (\step{4h}). \EDst uses the \texttt{get\_transport\_key} API to request \key from \SMDst (Steps~\circled{4j}-\circled{4k}), verifies $M$, and decrypts $C$ to recover \StateDst. However, \EDst does not process inputs until it receives a signal from \SMDst that the update completed successfully (cf. \step{4q}).

\vspace{0.5em}
\noindent\underline{Commit phase.}
After successfully restoring the state, \EDst uses the \texttt{migration\_commit} API (Ext. 11) to send a commit signal to \SMDst (\step{4m}). In response, \SMDst instructs \SMSrc to destroy \ESrc and to clear the corresponding migration metadata entry, finalizing the migration on the source device (\step{4n}).
\SMSrc then acknowledges successful destruction to \SMDst and sets another timeout $T$ (\step{4o}).
\SMDst clears its migration metadata 
and notifies both \EDst and \SMSrc that the migration was successful (Steps~\circled{4p}-\circled{4q}), such that \EDst starts processing inputs. 

If \SMSrc receives the acknowledgment in \step{4q} before $T$ expires, it notifies \party that migration completed successfully (\step{5}). 
Otherwise, it raises an alarm to \party for further investigation to determine whether the message was lost or the destination enclave failed to start. This behavior stems from intentionally destroying the source enclave before removing the lease at the destination, preventing both enclaves from running concurrently in the unlikely event of a timeout due to message loss. This design prioritizes security over automatic recovery by guaranteeing atomicity and state continuity (cf. \Cref{sec:security}). Prior to raising an alarm, \SMSrc may resend the message in \step{4o} a bounded number of times to complete the migration in the event of message loss. An alternative approach would defer destruction of the source enclave and its metadata until \SMSrc receives the acknowledgment in \step{4q}, enabling automatic recovery at the cost of a corner-case risk where both enclaves could be temporarily active if messages are lost.

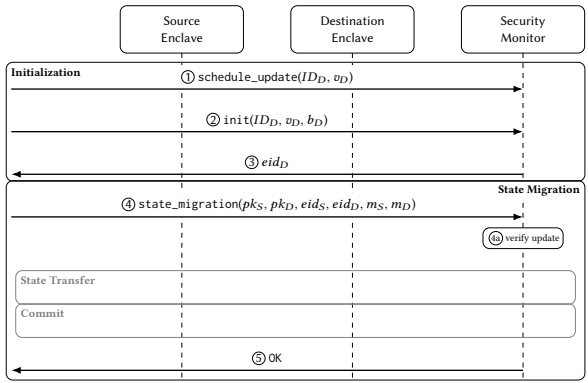
\begin{figure}
    \centering
    \scalebox{0.63}{

\begin{tikzpicture}[
  party/.style={draw, minimum width=2.6cm, minimum height=1cm, rounded corners},
  msg/.style={-{Latex[length=2mm]}, thick, draw=white},
  arrow/.style={-{Latex[length=2mm]}, thick},
  font=\sffamily\small,
  every node/.style={align=center},
  phase/.style={draw=black, rounded corners, inner sep=3pt, font=\bfseries\footnotesize},
  subphase/.style={draw=gray, rounded corners, inner sep=3pt, font=\bfseries\footnotesize}
]

\pgfdeclarelayer{bg}
\pgfdeclarelayer{fg} 
\pgfsetlayers{bg,main,fg}

\def\xHost{1}
\def\xSrc{4.6}
\def\xDst{8.2}
\def\xSM{11.8}
\def\yStep{0.9}
\def\nSteps{7.1}

\node[] (Host) at (\xHost, 0) {};
\node[party] (Src) at (\xSrc, 0) {Source\\Enclave};
\node[party] (Dst) at (\xDst, 0) {Destination\\Enclave};
\node[party] (SM) at (\xSM, 0) {Security\\Monitor};

\begin{pgfonlayer}{bg}
    \foreach \x in {\xSrc, \xDst, \xSM} {
      \draw[dashed] (\x, -0.5) -- ++(0, -\nSteps*\yStep - 0.5);
    }
\end{pgfonlayer}

\newcommand{\msg}[4]{%
  \path (#2) ++(0,-#1*\yStep) coordinate (from);
  \path (#3) ++(0,-#1*\yStep) coordinate (to);
  \draw[msg] (from) -- (to) node[midway, above, fill=white] {#4};
  \draw[arrow] (from) -- (to) {};
}

\newcommand{\event}[3]{%
  \path (#2) ++(0,-#1*\yStep - 0.5*\yStep) coordinate (pos);
  \node[fill=white, draw=black, rounded corners, font=\scriptsize, inner sep=2pt] at (pos) {#3};
}

\msg{1.4}{Host}{SM}{\stepCircled{1}\texttt{schedule\_update}(\IDDst, \vDst)}
\msg{2.4}{Host}{SM}{\stepCircled{2}\texttt{init}(\IDDst, \vDst, \binDst)}
\msg{3.4}{SM}{Host}{\stepCircled{3}\eidDst}
\msg{4.4}{Host}{SM}{~\stepCircled{4}\texttt{state\_migration}(\pkSrc, \pkDst, \eidSrc, \eidDst, \measSrc, \measDst)}
\event{4.4}{SM}{~\stepCircled{4a}verify update}
\msg{8.0}{SM}{Host}{\stepCircled{5}\texttt{OK}}

\node[phase, fit={(1,-0.8) (13,-3.1)}] (initbox) {};
\node[phase, fit={(1,-3.3) (13,-7.3)}] (migrationbox) {};
\node[subphase, fit={(1.2,-5.2) (12.8,-5.7)}] (transferbox) {};
\node[subphase, fit={(1.2,-5.9) (12.8,-6.4)}] (commitbox) {};

\begin{pgfonlayer}{bg}
    \node[anchor=north west, font=\bfseries\footnotesize] at (initbox.north west) {Initialization};
    \node[anchor=north east, font=\bfseries\footnotesize, fill=white] at (migrationbox.north east) {State Migration};
    \node[anchor=north west, font=\bfseries\footnotesize, fill=white] at (transferbox.north west) {\color{gray}State Transfer};
    \node[anchor=north west, font=\bfseries\footnotesize,fill= white] at (commitbox.north west) {\color{gray}Commit};
\end{pgfonlayer}

\end{tikzpicture} }
    \vspace{-0.2cm}
    \caption{Update Protocol, re-using the state transfer and commit phases from the migration protocol.}
    \vspace{-1.5 em}
    \label{fig:update_protocol}
\end{figure}

\subsection{Secure Enclave Update}
\label{sec:update}

We now describe the update protocol of our toolkit.
Conceptually, an update is a special case of a migration: instead of transferring state to an enclave with identical software on another device, the state is migrated to an enclave with updated software on the same device. Consequently, it follows the same sequence of steps as the migration protocol, but with a different initialization phase.

At the start of an update, a source enclave \ESrc is executing the binary \binSrc with version \vSrc for software \IDSrc. Its current state is denoted \StateSrc. The update is initiated by a party \party, which may be local or remote. During the update, a new enclave \EDst is created from the updated binary $b_D$ with version \vDst and the same software identifier: $ID_D=ID_S$. The protocol ensures that the state is securely transferred, i.e., $D_D=D_S$, to a new version of the software ($\vDst>\vSrc$) and that \ESrc and \EDst are never executed concurrently.

Updates may be triggered by security patches, feature extensions, or policy-driven lifecycle events. Upon such an event, $P$ initiates a code update, sending the software ID \IDDst and new version \vDst to the SM via the \texttt{schedule\_update} API (Ext. 6, \step{1}). Similar to the migration protocol, \SM verifies that \party is authorized to initiate an update. It additionally validates that an enclave with the corresponding software $\IDSrc=\IDDst$ exists before scheduling an update and creating a new (inactive) enclave \EDst from \binDst upon initialization request from \party (\step{2}). 

The update protocol then proceeds with the state migration detailed in \Cref{sec:migration}, applying the following minor modifications.
While the state migration request of the migration protocol includes equal measurements for source and destination enclave, the update's state migration request comprises equal public keys ($pk_S = pk_D$) (\step{4}). When detecting this, \SM performs a different request verification (\step{4a}), confirming that \IDDst matches \IDSrc and $v_D > v_S$ instead of verifying that $\measSrc=\measDst$ (\cf Mod. 2).
Since an update is executed on a single SM, no inter-SM synchronization is required. Consequently, several protocol steps are handled internally: \SM can start the destination enclave directly, merging Steps~\circled{4f} and~\circled{4g} into a single step; and it can destroy the source enclave immediately without confirming this action, combining Steps~\circled{4n} and~\circled{4o}, and omitting \step{4q}.

A full summary of the update protocol is shown in \Cref{fig:update_protocol}.

\section{Security Analysis}
\label{sec:security}

We now analyze the security of \lib (cf. \Cref{sec:requirements}).

\subsection{Properties S1---S4}
Note that four security properties follow directly from our design:
\begin{description}[leftmargin=0.2cm]
    \item[S1---Authenticity.] 
    In both protocols, the initiating party sends an initialization message, containing the public keys of the source and destination SMs, thereby establishing trusted SMs (\cf\step{4}). The messages are transmitted over an authenticated channel secured with a MAC, guaranteeing both message origin authenticity and integrity.

    \item[S2---Code Integrity.] 
    All binaries are transmitted to the destination SM via a secure, MAC-authenticated channel (\cf\step{2}). Any modification is detected before the binary is instantiated. Furthermore, the derivation function for the ephemeral state encryption key ($\key=H(\seed,\measSrc,\measDst)$) involves the enclave measurements, further enforcing that the state can only be imported if the source and destination measurements match the specified values.

    \item[S3---State Confidentiality and Integrity.] 
    In the migration and update protocols, the state leaves enclave protection only in \step{4e}, encrypted under an ephemeral symmetric key and authenticated with a MAC. This ensures both confidentiality and integrity of the transferred state.  
    
    \item[S4---Software Rollback Protection.] 
    To prevent downgrading to prior software versions, the SM enforces monotonic versioning. The key generation (\cf\step{4c}) binds the state decryption key to the enclave measurements. In a migration, this ensures that only a destination enclave with $\measDst=\measSrc$ can decrypt the state, preventing rollback to older versions. 
    In updates, the SM verifies that $v_D > v_S$ (\cf\step{4a}), likewise preventing rollback to older versions. Once a newer version is recorded, older ones are rejected. Version information is securely stored in the SM’s encrypted, integrity-protected, rollback-resistant non-volatile memory (\cf\Cref{sec:sm_extensions}). 
    If rollback-resistant memory is unavailable, rollback protection requires a monotonic counter service~\cite{DBLP:conf/uss/MateticAKDSGJC17/ROTE,DBLP:conf/ccs/NiuPZZ22/Narrator}. Without such support, the SM itself could be rolled back, allowing an adversary to restore an old state for a specific enclave. In this case, software rollback protection is effective only while version information resides in the SM’s runtime memory. 
\end{description} 

\noindent
We now analyze the two remaining properties in more detail.

\subsection{S5---Atomicity}
\label{sec:security_atomicity}

There are four possible cases during migration that could affect atomicity. We discuss them case by case and show how our design preserves atomicity in all situations.

\begin{description}[leftmargin=0.2cm]
    \item[SM failure:] If any SM operation (e.g., key generation or destroying \ESrc) fails due to faults or interference, the SM treats the operation as failed. It then: (1) destroys \EDst and blacklists \measDst (if it is \SMDst), (2) resumes \ESrc (if it is \SMSrc), (3) clears migration metadata, and (4) notifies \party and the remote SM. The remote SM mirrors these steps to ensure only \ESrc remains active.

    \item[SM crash:] If an SM crashes before \step{4o} completes, it restores migration metadata from persistent memory and triggers the same recovery steps as above. Only \SMSrc allows \ESrc to resume after recovery, while \SMDst prevents resumption of \EDst.
    As with S4, atomicity during a crash depends on rollback-protected persistent memory; without it, an adversary could replay stale migration metadata, preventing \EDst from being destroyed or \ESrc from resuming, and leaving either both enclaves running or neither running after a failed migration. 
    
    \item[Lost or delayed messages:] Most protocol messages are sent over authenticated and encrypted channels and cannot be forged. State export/import requests (Steps~\circled{4b} and~\circled{4h} are unauthenticated but can only succeed if tied to an active migration in the SM. Invalid or out-of-sequence requests are rejected. Moreover, delays or losses cause $T_{SM}$ to expire, triggering the same recovery steps as in the first scenario.  
    A special case occurs if \step{4o} fails: \SMDst's timeout expires and neither enclave is running. In this case, \SMSrc detects the anomaly when its timeout $T$ expires before receiving message \circled{4q}, allowing it to raise an alarm to \party. 
    
    \item[Enclave crashes:] If \ESrc or \EDst crashes (e.g., before sending the commit in \step{4m}), the enclave can be resumed and the failed step retried. The SM enforces ordering: \EDst can only be activated after \step{4g}, and \ESrc only before that step. If the enclave is not resumed in time, $T_{SM}$ expires and the SM reverts to a safe state as in the previous scenarios with only \ESrc active. 
\end{description}

If none of these scenarios occur, migration proceeds to \step{4n}, where \SM destroys \ESrc, keeping only \EDst active. Only then does \step{4q} allow \EDst to process inputs. As Steps~\circled{4f}-\circled{4g} ensure \EDst is only activated after \ESrc is paused, the two enclaves never execute concurrently.  Together, these additional steps enforce atomicity: at any time, either \ESrc or \EDst is active, but never both.

The update protocol follows a similar structure but involves only one SM, mirroring the migration steps. Removing synchronization needs does not violate atomicity, as the SM can directly perform the corresponding steps (e.g., destruction of \ESrc) internally, removing the risk of lost messages or crashes.

\subsection{S6---State Continuity}

State continuity is preserved in \lib because the source (\ESrc) and destination (\EDst) enclaves (1) never execute concurrently and (2) execute atomically. Specifically, \ESrc halts input processing once it exports its state in \step{4e}, and \EDst only begins processing inputs after the SM signals migration completion in \step{4p}, at which point \ESrc has already been destroyed. This strictly sequential handover ensures that no two enclave versions process distinct inputs in parallel, thereby preventing state divergence. The same reasoning applies to updates.
Furthermore, the exported state is encrypted using an \emph{ephemeral} session key known only to the involved enclaves and SM(s). This guarantees that state material from failed or aborted updates and migrations cannot be reused or decrypted by an adversary, eliminating replay-based state forking.

Our cloning protection mechanism ensures that no additional enclaves with the same software identifier are instantiated during updates or migrations, preventing concurrent execution of identical enclaves that could violate state continuity (\cf\Cref{sec:rollback_cloning}).
When multiple enclaves are genuinely allowed to execute, \lib distinguishes between independent clones (with isolated state) and state-sharing clones (with synchronized state). For state-sharing clones, all but one instance must be terminated before an update or migration to preserve consistency. Independent clones may proceed concurrently, but all instances must complete the update or migration together.
The SM enforces these guarantees by (i) allowing execution switches and source enclave destruction only after all participating clones have requested the update key and confirmed their commit, (ii) ensuring each enclave identifier participates in only one update or migration session, and (iii) propagating the upper bound $N$ on clones to all destination enclaves. Each clone receives a unique ephemeral encryption key, restricting state decryption to its designated destination. If multiple clones are active and the clone type is unspecified, the SM aborts the operation.

Aside from update and migration events, state continuity is enforced by the state continuity module (\cf\Cref{sec:rollback_cloning}). The SM enforces the enclave-specified limit $N$ on active instances, preventing unauthorized clones from manipulating shared state. It also blocks initialization of outdated enclave versions to prevent execution on stale code or data.
Finally, to protect against rollback-based state reversion, \lib provides software-based monotonic counters. Enclaves can embed these counters in their sealed state as freshness tokens, ensuring that any replayed or stale state can be detected upon restoration. This mechanism guarantees that enclaves always resume execution from the most recent, consistent state. However, maintaining consistency across SM restarts requires that these monotonic counters themselves be protected against rollback; e.g., by storing them in rollback-resistant persistent memory.

\section{Implementation \& Evaluation}
\label{sec:performance}

We now evaluate the performance of \lib integrated in Keystone against the requirements outlined in \Cref{sec:requirements}.

\newcommand{\memRow}[7]{\textbf{#2} & #1 & #3 & #4 & #5 & #6 & #7 \\ }

\begin{table}[tbp]
    \centering
    \footnotesize
    \caption{Memory overhead introduced by the additional data structures of the Trusted Time (TT), State Continuity (SC), and Update/Migration (UM) modules. Within the SC module, we further distinguish memory used for software rollback (SW), cloning (CL), and rollback (RB) protection.}

    \scalebox{0.92}{
    \begin{tabular}{llccccc}
        \toprule
         & & \multicolumn{5}{c}{\textbf{size (B) / entry}} \\ \cmidrule{3-7}
         & & \textbf{TT} & \multicolumn{3}{c}{\textbf{SC}} & \textbf{UM} \\ \cmidrule{4-6}
         \textbf{Component} & \textbf{Data structure} & & \textbf{SW} & \textbf{CL} & \textbf{RB} &  \\ 
         \midrule

        \memRow{Enclave metadata}{Mod. 1}{16}{16}{8}{0}{4}

        \memRow{SW version tracking}{Ext. 1}{0}{16}{0}{0}{0}
        \memRow{Monotonic counters}{Ext. 2}{0}{0}{0}{24}{0}
        \memRow{SW $ID$s for update/migration}{Ext. 3}{0}{0}{0}{0}{8}
        \memRow{Migration metadata}{Ext. 4}{0}{0}{0}{0}{224}

        \bottomrule
    \end{tabular}}
    \label{tab:mem_overhead}
\end{table}

\subsection{Experimental Setup}

We implemented a prototype of \lib by extending the Keystone SM with modules for trusted time, state continuity, update, and migration support. The extensions were integrated directly into the SM source and compiled into its binary. Functions for monotonic counter management, enclave-local time retrieval, and transport key access are exposed to enclaves via SBI calls, while sensitive operations (e.g., update scheduling and execution switch) are protected by a dedicated \texttt{migration\_message} interface, which encrypts and authenticates protocol messages exchanged between SMs and the device operator. This design choice minimizes the TCB and prevents the host from inferring enclave update or migration states. The Keystone SDK was also extended to expose a high-level API for easy access to these functions, similar to a wrapping library for developers (\cf\Cref{sec:system_model}).
Our implementation comprises approximately 750 lines of code (LoC): 50 for trusted time, 160 for state continuity, and 540 for update and migration support.

Our evaluation does not include TLS session management within the SM. Instead, we assume that TLS sessions between the owner and the enclave, as well as between the two security monitors are established in advance to pre-share session keys used for encryption and MAC generation. We stress that excluding TLS session management does not affect our evaluation results. Namely, prior work demonstrates that a full TLS handshake on constrained IoT devices typically introduces approximately 100–200 ms of additional latency (e.g., around 147 ms in representative measurements)~\cite{tlsHandshakeLatency}), whereas our update protocol requires up to 3.5 s for 1 MB states (cf. Figure~\ref{fig:bench_update_migration}).
All measurements were performed on a VisionFive2 board equipped with a StarFive JH7110 (RISC-V U74 quad-core, 1.5 GHz, 8 GB RAM). We evaluated each module under both ideal conditions (no background processes) and with considerable processes running in the background. We emulated such background load using the \texttt{stress-ng} tool for CPU-intensive (\verb|--cpu 4|), memory-intensive (\verb|--vm 2 --vm-bytes 70%|), and mixed workloads. Note that evaluating \lib under more realistic workloads such as the Phoronix Benchmark Suite was not possible, as the Buildroot-generated image for the VisionFive2 board lacks a package manager and provides only minimal tooling. For the evaluation of the update, migration, and state continuity modules, state sizes ranged from 16 B to 1 MB. Unless specified otherwise, each data point in our measurements is averaged over 10 independent runs.

\subsection{Evaluation Results}

\subheading{Update and Migration.} 
We evaluated update and migration latency using paired source and destination applications instrumented with \lib’s API. Updates were performed locally on one board, while migrations used two identical VisionFive2 boards connected via WAN over reverse SSH, with a Python controller emulating \party. 
The network exhibited an average bandwidth of 311 Mbit/s and latencies ranging from 30.053 ms to 30.28 ms (mean 30.16 ms, standard deviation 0.06 ms). Note that updates are largely insensitive to network variability, as messages exchanged over the network are fixed in size and capped at 144 B, except when transmitting the updated software to the target platform (which is a process that does not contribute to service downtime).

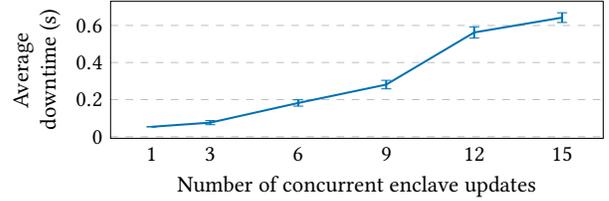
\begin{figure}[tbp]
    \centering
    \scalebox{0.95}{
        \begin{tikzpicture}

\begin{axis}[
    width=1.0\columnwidth,
    height=3.5cm,
    log basis y=2,
    xlabel={Number of concurrent enclave updates},
    ylabel={\parbox{2cm}{\centering Average \\ downtime (s)}},
    legend style={
        at={(0.5,1.05)},
        anchor=south,
        legend columns=-1,
        font=\small,
        /tikz/every even column/.append style={column sep=0.5cm}
    },
    grid=both,
    minor grid style={dotted},
    major grid style={dashed},
    xmajorgrids=false,
    xminorgrids=false,
    ymajorgrids=true,
    yminorgrids=false,
    tick style={draw=none}, 
    xtick={1,3,6,9,12,15},
    xticklabels={
        1,
        3,
        6,
        9,
        12,
        15,
    },
]

\addplot[
    color={rgb,255:red,0;green,114;blue,178},
    thick,
    solid,
    error bars/.cd,
        y dir=both,
        y explicit,
        error bar style={line width=0.8pt},
        error mark options={rotate=90, mark size=2pt}
] coordinates {
    (1, 0.0532) +- (0, 0.0006)
    (3, 0.0756) +- (0, 0.0110)
    (6, 0.1817) +- (0, 0.0170)
    (9, 0.2809) +- (0, 0.0230)
    (12, 0.5622) +- (0, 0.0304)
    (15, 0.6423) +- (0, 0.0263)
};

\end{axis}

\end{tikzpicture}
    }
    \vspace{-0.3cm}
    \caption{Impact of concurrent updates on service downtime; here, we assume a 1 KB state.}
    \label{fig:bench_concurrency}
    \vspace{-1 em}
\end{figure}

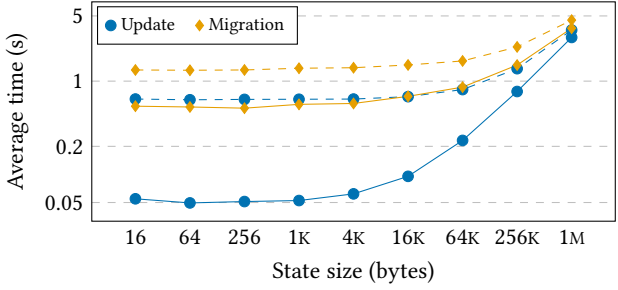
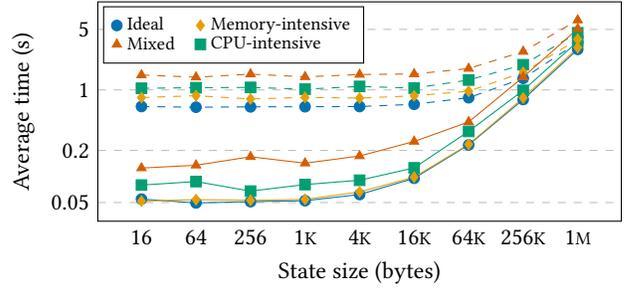
\begin{figure*}[tbp]
    \centering
    \begin{subfigure}{.48\textwidth}
        \centering
            \begin{tikzpicture}

\begin{axis}[
    width=1.0\textwidth,
    height=4.5cm,
    xmode=log,
    log basis x=2,
    ymode=log,
    log basis y=2,
    xlabel={State size (bytes)},
    ylabel={Average time (s)},
    legend style={
        at={(0.01,0.98)},
        anchor=north west,
        legend columns=-1,
        font=\small,
        /tikz/every even column/.append style={column sep=0.2cm}
    },
    grid=both,
    minor grid style={dotted},
    major grid style={dashed},
    xmajorgrids=false,
    xminorgrids=false,
    ymajorgrids=true,
    yminorgrids=false,
    tick style={draw=none}, 
    ytick={0.05,0.2,1,5},
    yticklabels={0.05,0.2,1,5},
    xtick={16,64,256,1024,4096,16384,65536,262144,1048576},
    xticklabels={
        16,
        64,
        256,
        {$1\text{\footnotesize K}$},
        {$4\text{\footnotesize K}$},
        {$16\text{\footnotesize K}$},
        {$64\text{\footnotesize K}$},
        {$256\text{\footnotesize K}$},
        {$1\text{\footnotesize M}$}
    },
]

\addlegendimage{only marks, color={rgb,255:red,0;green,114;blue,178}, mark=*, mark options={solid, fill={rgb,255:red,0;green,114;blue,178}}}
\addlegendentry{\footnotesize Update}

\addlegendimage{only marks, color={rgb,255:red,230;green,159;blue,0}, mark=diamond*, mark options={solid, fill={rgb,255:red,230;green,159;blue,0}}}
\addlegendentry{\footnotesize Migration}

\addplot+[
    color={rgb,255:red,0;green,114;blue,178},
    mark=*,
    solid,
    mark options={solid, fill={rgb,255:red,0;green,114;blue,178}}
] coordinates {
    (16, 0.0548)
    (64, 0.0495)
    (256, 0.0512)
    (1024, 0.0525)
    (4096, 0.0619)
    (16384, 0.0954)
    (65536, 0.2315)
    (262144, 0.7743)
    (1048576, 2.9406)
};

\addplot+[
    color={rgb,255:red,0;green,114;blue,178},
    mark=*,
    dashed,
    mark options={solid, fill={rgb,255:red,0;green,114;blue,178}}
] coordinates {
    (16, 0.6429)
    (64, 0.6313)
    (256, 0.6371)
    (1024, 0.6394)
    (4096, 0.6426)
    (16384, 0.6817)
    (65536, 0.8122)
    (262144, 1.3599)
    (1048576, 3.5304)
};

\addplot+[
    color={rgb,255:red,230;green,159;blue,0},
    mark=diamond*,
    solid,
    mark options={solid, fill={rgb,255:red,230;green,159;blue,0}}
] coordinates {
    (16, 0.5388)
    (64, 0.5285)
    (256, 0.5131)
    (1024, 0.5639)
    (4096, 0.5762)
    (16384, 0.6872)
    (65536, 0.8631)
    (262144, 1.4874)
    (1048576, 3.6918)
};

\addplot+[
    color={rgb,255:red,230;green,159;blue,0},
    mark=diamond*,
    dashed,
    mark options={solid, fill={rgb,255:red,230;green,159;blue,0}}
] coordinates {
    (16, 1.3175)
    (64, 1.3081)
    (256, 1.3179)
    (1024, 1.3749)
    (4096, 1.3925)
    (16384, 1.4906)
    (65536, 1.6457)
    (262144, 2.327)
    (1048576, 4.5076)
};

\end{axis}

\end{tikzpicture}
        \label{fig:bench_migration}
        \vspace{-0.9cm}
        \caption{Update and migration latency.}
    \end{subfigure}
    \hfill
    \begin{subfigure}{.48\textwidth}
        \centering
            \begin{tikzpicture}
	
\begin{axis}[
    width=1.0\textwidth,
    height=4.5cm,
    xmode=log,
    log basis x=2,
    ymode=log,
    log basis y=2,
    xlabel={State size (bytes)},
    ylabel={Average time (s)},
    legend style={
        at={(0.01,0.98)},
        anchor=north west,
        legend columns=2,
        font=\small,
        row sep=-3pt,
        cells={anchor=west},
        /tikz/every even column/.append style={column sep=0.2cm}
    },
    grid=both,
    minor grid style={dotted},
    major grid style={dashed},
    xmajorgrids=false,
    xminorgrids=false,
    ymajorgrids=true,
    yminorgrids=false,
    tick style={draw=none}, 
    ytick={0.05,0.2,1,5},
    yticklabels={0.05,0.2,1,5},
    xtick={16,64,256,1024,4096,16384,65536,262144,1048576},
    xticklabels={
        16,
        64,
        256,
        {$1\text{\footnotesize K}$},
        {$4\text{\footnotesize K}$},
        {$16\text{\footnotesize K}$},
        {$64\text{\footnotesize K}$},
        {$256\text{\footnotesize K}$},
        {$1\text{\footnotesize M}$}
    },
]

\addlegendimage{only marks, color={rgb,255:red,0;green,114;blue,178}, mark=*, mark options={solid, fill={rgb,255:red,0;green,114;blue,178}}}
\addlegendentry{\footnotesize Ideal}

\addlegendimage{only marks, color={rgb,255:red,230;green,159;blue,0}, mark=diamond*, mark options={solid, fill={rgb,255:red,230;green,159;blue,0}}}
\addlegendentry{\footnotesize Memory-intensive}

\addlegendimage{only marks, color={rgb,255:red,213;green,94;blue,0}, mark=triangle*, mark options={solid, fill={rgb,255:red,213;green,94;blue,0}}}
\addlegendentry{\footnotesize Mixed}

\addlegendimage{only marks, color={rgb,255:red,0;green,158;blue,115}, mark=square*, mark options={solid, fill={rgb,255:red,0;green,158;blue,115}}}
\addlegendentry{\footnotesize CPU-intensive}

\addplot+[
    color={rgb,255:red,0;green,114;blue,178},
    mark=*,
    solid,
    mark options={solid, fill={rgb,255:red,0;green,114;blue,178}}
] coordinates {
    (16, 0.0548)
    (64, 0.0495)
    (256, 0.0512)
    (1024, 0.0525)
    (4096, 0.0619)
    (16384, 0.0954)
    (65536, 0.2315)
    (262144, 0.7743)
    (1048576, 2.9406)
};

\addplot+[
    color={rgb,255:red,0;green,114;blue,178},
    mark=*,
    dashed,
    mark options={solid, fill={rgb,255:red,0;green,114;blue,178}}
] coordinates {
    (16, 0.6429)
    (64, 0.6313)
    (256, 0.6371)
    (1024, 0.6394)
    (4096, 0.6426)
    (16384, 0.6817)
    (65536, 0.8122)
    (262144, 1.3599)
    (1048576, 3.5304)
};

\addplot+[
    color={rgb,255:red,0;green,158;blue,115},
    mark=square*,
    solid,
    mark options={solid, fill={rgb,255:red,0;green,158;blue,115}}
] coordinates {
    (16, 0.0796)
    (64, 0.0874)
    (256, 0.0678)
    (1024, 0.0807)
    (4096, 0.0905)
    (16384, 0.1257)
    (65536, 0.3319)
    (262144, 0.9873)
    (1048576, 3.6289)
};

\addplot+[
    color={rgb,255:red,0;green,158;blue,115},
    mark=square*,
    dashed,
    mark options={solid, fill={rgb,255:red,0;green,158;blue,115}}
] coordinates {
    (16, 1.0444)
    (64, 1.0606)
    (256, 1.0686)
    (1024, 1.0193)
    (4096, 1.0969)
    (16384, 1.0491)
    (65536, 1.3011)
    (262144, 1.9544)
    (1048576, 4.5705)
};

\addplot+[
    color={rgb,255:red,230;green,159;blue,0},
    mark=diamond*,
    solid,
    mark options={solid, fill={rgb,255:red,230;green,159;blue,0}}
] coordinates {
    (16, 0.0519)
    (64, 0.054)
    (256, 0.053)
    (1024, 0.0544)
    (4096, 0.0664)
    (16384, 0.0983)
    (65536, 0.2354)
    (262144, 0.8067)
    (1048576, 3.0786)
};

\addplot+[
    color={rgb,255:red,230;green,159;blue,0},
    mark=diamond*,
    dashed,
    mark options={solid, fill={rgb,255:red,230;green,159;blue,0}}
] coordinates {
    (16, 0.8155)
    (64, 0.8573)
    (256, 0.786)
    (1024, 0.8197)
    (4096, 0.8043)
    (16384, 0.8545)
    (65536, 0.9657)
    (262144, 1.5953)
    (1048576, 3.8394)
};

\addplot+[
    color={rgb,255:red,213;green,94;blue,0},
    mark=triangle*,
    solid,
    mark options={solid, fill={rgb,255:red,213;green,94;blue,0}}
] coordinates {
    (16, 0.1252)
    (64, 0.1348)
    (256, 0.169)
    (1024, 0.1424)
    (4096, 0.1725)
    (16384, 0.2537)
    (65536, 0.4266)
    (262144, 1.4063)
    (1048576, 5.0775)
};

\addplot+[
    color={rgb,255:red,213;green,94;blue,0},
    mark=triangle*,
    dashed,
    mark options={solid, fill={rgb,255:red,213;green,94;blue,0}}
] coordinates {
    (16, 1.4866)
    (64, 1.4031)
    (256, 1.5216)
    (1024, 1.4153)
    (4096, 1.506)
    (16384, 1.5278)
    (65536, 1.7651)
    (262144, 2.7758)
    (1048576, 6.3994)
};

\end{axis}

\end{tikzpicture}
        \label{fig:bench_update}
        \vspace{-0.9cm}
        \caption{Update latency with varying background workloads.}
    \end{subfigure}
    \vspace{-0.3cm}
    \caption{Latency of the enclave update and migration protocols for varying state sizes. We report the end-to-end update/migration time (dashed lines) and the time span where the enclave service is unavailable (solid lines).}
    \label{fig:bench_update_migration}
\end{figure*}

\Cref{fig:bench_update_migration} shows the results. Overall, the update protocol exhibits low latency and high robustness across all conditions. In the absence of background load, the end-to-end update time remains stable around 0.6-0.7 s for states up to 16 KB, while service downtime remains below as little as 10 ms. Both metrics scale linearly with the state size, primarily due to the cost of encrypting and decrypting state during the state transfer phase. For the largest state (1 MB), the total update time reaches 3.5 s, with a service downtime of 2.9~s.
Due to the limited number of PMP registers, the VisionFive2 board supports at most three concurrent enclaves. Therefore, to emulate concurrent updates in our testbed, we interleaved multiple update requests per enclave (and streamlined destination initialization, source destruction, and a few SM checks during updates).\footnote{Our benchmarks confirm that these simplifications do not affect service downtime.} \Cref{fig:bench_concurrency} shows the latency introduced by up to 15 concurrent enclave updates, which reaches up to {\color{black}0.64 seconds of downtime} under high concurrency load, with a maximum end-to-end latency of a few seconds.

The migration protocol exhibits increased latency compared to updates, dominated by inter-device communication and state transfer over the network. The migration time remains around 1.3-1.5 s for state migrations up to 16 KB, maintaining a service downtime below 700 ms. Similar to the update protocol, migration time scales linearly with the state size, additionally including state transfer between the target devices. Total migration time reaches a maximum of 4.5 s end-to-end latency and 3.7 s downtime for the largest state. 
This overhead remains acceptable for most applications, especially in automotive scenarios where updates and migrations are not expected to occur while the vehicle is in operation.

\begin{figure}[t]
	\centering
	\begin{minipage}[b]{0.49\linewidth} 
		\centering
		\vspace{-1 em}
		\scalebox{0.9}{
			\begin{tikzpicture}[baseline=(current bounding box.south)]

\definecolor{oipBlue}{RGB}{0,114,178}
\definecolor{oipOrange}{RGB}{230,159,0}
\definecolor{oipGreen}{RGB}{0,158,115}
\definecolor{oipRed}{RGB}{213,94,0}

\begin{axis}[
    ybar,
    bar width=0.3cm,
    width=1.0\textwidth,
    height=4.0cm,
    ylabel={Average time ($\mu$s)},
    ymin=0,
    xtick={1,2},
    xticklabels={
        {$\text{E-H}$},
        {$\text{H-E}$}
    },
    x=1.3cm,
    enlarge x limits={abs=0.5cm},
    grid=both,
    minor grid style={dotted},
    major grid style={dashed},
    xmajorgrids=false,
    xminorgrids=false,
    ymajorgrids=true,
    yminorgrids=false,
    tick style={draw=none},
    name=myaxis,
]

\addplot[fill=oipBlue!90, draw=oipBlue] coordinates {
    (1, 2.92)
    (2, 3.05)
};

\addplot[fill=oipRed!90, draw=oipRed] coordinates {
    (1, 2.94)
    (2, 3.50)
};

\end{axis}

\node[
    anchor=center,
    rotate=90,
    font=\small,
    draw,
    inner sep=2pt,
] at ([xshift=0.5cm] myaxis.east) {%
    \setlength{\tabcolsep}{0pt}%
    \begin{tabular}{l}
        \raisebox{1pt}[\height][1pt]{%
            \tikz[baseline=0.05cm]{\draw[fill=oipBlue!90, draw=oipBlue] (0,0) rectangle (0.3,0.2);
                  \node[anchor=west] at (0.35,0.1) {\strut enabled};}%
        } \\[-2pt]
        \raisebox{1pt}[\height][1pt]{%
            \tikz[baseline=0.05cm]{\draw[fill=oipRed!90, draw=oipRed] (0,0) rectangle (0.3,0.2);
                  \node[anchor=west] at (0.35,0.1) {\strut disabled};}%
        }
    \end{tabular}
};

\end{tikzpicture}
		}
		\caption{Overhead for local-time tracking during context switches from enclave to host (E-H) and vice versa (H-E).}
		\label{fig:timer_ctx_switch}
	\end{minipage}
	\hfill
	\begin{minipage}[b]{0.49\linewidth}
		\centering
		\scalebox{0.9}{
			\begin{tikzpicture}[baseline=(current bounding box.south)]

\begin{axis}[
    ybar,
    bar width=0.4cm,
    width=1.0\columnwidth,
    height=4.0cm,
    ylabel={Average time ($\mu$s)},
    ymin=0,
    xtick={1,2,3},
    xticklabels={
        rdtime,
        SM glob,
        SM loc
    },
    x=1.2cm,
    enlarge x limits={abs=0.5cm}, 
    grid=both,
    minor grid style={dotted},
    major grid style={dashed},
    xmajorgrids=false,
    xminorgrids=false,
    ymajorgrids=true,
    yminorgrids=false,
    tick style={draw=none},
]

\definecolor{oipBlue}{RGB}{0,114,178}

\addplot[fill=oipBlue!90, draw=oipBlue] coordinates {
    (1, 0.36)
    (2, 7.18)
    (3, 7.22)
};

\end{axis}

\end{tikzpicture}
		}
		\caption{Latency of consecutive time queries using \texttt{rdtime} and SM interfaces for global and enclave-local time.}
		\label{fig:timer_read}
	\end{minipage}
	\begin{minipage}[b]{0.99\linewidth}
		\centering
		\vspace{1.0 em}
		\scalebox{0.9}{
			\begin{tikzpicture}
	
\begin{axis}[
    ybar=1pt,
    bar width=0.15cm,
    width=0.45\textwidth,
    height=4.0cm,
    ymode=log,
    log basis y=2,
    ylabel={Average time (ms)},
    xlabel={State size (bytes) },
    ymin=1.5,
    ymax=2000,
    ytick={2, 10, 100, 1000},
    yticklabels={2,10,100,1000},
    xtick={1,2,3,4,5,6,7,8, 9},
    xticklabels={
        16,
        64,
        256,
        {$1\text{\footnotesize K}$},
        {$4\text{\footnotesize K}$},
        {$16\text{\footnotesize K}$},
        {$64\text{\footnotesize K}$},
        {$256\text{\footnotesize K}$},
        {$1\text{\footnotesize M}$}
    },
    x=0.8cm,
    enlarge x limits={abs=0.5cm},
    legend style={at={(0.02,0.98)}, anchor=north west, legend columns=2},
    grid=both,
    minor grid style={dotted},
    major grid style={dashed},
    xmajorgrids=false,
    xminorgrids=false,
    ymajorgrids=true,
    yminorgrids=false,
    tick style={draw=none},
    legend entries={Read, Read (MC), Write, Write (MC)},
]

\definecolor{oipBlue}{RGB}{0,114,178}
\definecolor{oipRed}{RGB}{213,94,0}

\addplot[fill=oipBlue!90, draw=oipBlue] coordinates {
    (1, 2.2806)
    (2, 2.3883)
    (3, 2.5819)
    (4, 3.6564)
    (5, 7.8837)
    (6, 24.4323)
    (7, 91.0815)
    (8, 357.7343)
    (9, 1424.0642)
};

\addplot[fill=oipBlue!90, draw=oipBlue, postaction={pattern=north east lines}] coordinates {
    (1, 2.3029)
    (2, 2.3716)
    (3, 2.6744)
    (4, 3.6831)
    (5, 7.9136)
    (6, 24.5276)
    (7, 91.1082)
    (8, 357.7624)
    (9, 1424.1518)
};

\addplot[fill=oipRed!90, draw=oipRed] coordinates {
    (1, 2.4632)
    (2, 2.5452)
    (3, 2.7528)
    (4, 3.8233)
    (5, 12.3857)
    (6, 24.6975)
    (7, 91.7158)
    (8, 359.1181)
    (9, 1489.5852)
};

\addplot[fill=oipRed!90, draw=oipRed, postaction={pattern=north east lines}] coordinates {
    (1, 2.4963)
    (2, 2.5678)
    (3, 2.8666)
    (4, 3.8605)
    (5, 10.872)
    (6, 24.8016)
    (7, 91.5044)
    (8, 359.8803)
    (9, 1489.4856)
};

\end{axis}

\end{tikzpicture}
		}
		\vspace{-0.3cm}
		\caption{Overhead of using monotonic counters during state read and write operations for different state sizes.}
		\vspace{-2 em}
		\label{fig:mc_overhead}
	\end{minipage}
\end{figure}

Under CPU- or memory-intensive background workloads, both update and migration latencies increase moderately, with CPU-bound workloads having a greater impact than those cases where memory is in high contention. The combined workload of both CPU and memory-intensive tasks
shows the highest overhead, reaching 6.4 s end-to-end latency for updates and migrations for large states, but preserving similar linear scaling.
Furthermore, the module introduces only a modest memory overhead in the SM---4 B per enclave to indicate execution permission, 8 B for each scheduled update or migration, and 224 B of metadata per active operation.

Based on our measurements, we conclude that state size is the primary factor influencing service downtime. For state sizes up to 16 KB, updates incur end-to-end latency below 0.7 s with downtime under 10 ms, while migrations complete within 1.5 s with downtime below 700 ms. Beyond this range, downtime increases to several seconds. We therefore identify the practical operating envelope of the system as enclave states in the low tens of kilobytes when sub-second service interruption is required. For latency-sensitive services, we recommend limiting migratable state to less than 4 KB.

\vspace{0.25cm}
\subheading{Trusted Time.} 
We also measured the impact of the trusted time module on context switching and timestamp queries, averaging over 100 operations. The additional overhead for enclave-to-host and host-to-enclave context switches was 0.02 $\mu$s and 0.45 $\mu$s, re-spectively---negligible in practice (\cf\Cref{fig:timer_ctx_switch}).

Querying enclave-local time via the SM incurs a latency of 7.22 $\mu$s, compared to 0.36 $\mu$s for direct \texttt{rdtime} access (\cf\Cref{fig:timer_read}). This overhead is mainly due to enclave-SM context switching rather than timekeeping logic, as evidenced by indirect \texttt{rdtime} measurements from the SM. 
Enclave-local time is primarily used for non-time-critical tasks (e.g., lease validation or subscription accounting), where microsecond-level overhead is acceptable. Time-critical logic, such as speed or distance calculations, relies on global time, which remains accessible inside the enclave via the low-latency \texttt{rdtime} instruction.  
The module's memory footprint is minimal, adding only 16 B per enclave for time metadata (\cf\Cref{tab:mem_overhead}).

\vspace{0.25cm}
\subheading{State continuity.}
Most components of the state continuity module execute only once per enclave lifecycle and thus incur negligible runtime cost.
Both cloning and software rollback prevention require a check by the SM during enclave creation, while an enclave needs to allocate and free a monotonic counter once. We therefore focus on the frequently used \texttt{inc\_MC} and \texttt{get\_MC\_value} APIs, required for writing and reading state from disk, respectively.
Including the monotonic counter value in state sealing operations adds 0.88–1.63\% latency for small (16 B) states and below 0.01\% for 1 MB states (\Cref{fig:mc_overhead}). Memory overhead is similarly small: 24 B per counter in the SM, 8 B in sealed state, 8 B for cloning metadata, and 16 B per software ID for rollback protection (\cf\Cref{tab:mem_overhead}).

\section{Related Work}

To the best of our knowledge, no prior work provides a comprehensive toolkit addressing trusted time, state continuity, or secure, state-preserving updates and migrations for RISC-V TEEs. We therefore review related efforts for TEEs based on other ISAs.

\vspace{0.2cm}
\subheading{Trusted time.}
Support for trusted time varies across TEEs~\cite{alder2023aboutTime}. Intel TDX includes a trusted timer, but SGX removed this functionality in 2020. In Arm TrustZone, timekeeping is implementation-specific and not part of the core architecture. Crucially, no related work addresses support for enclave-local trusted time.

\vspace{0.2cm}
\subheading{State continuity.}
TEE cloning has been extensively studied for SGX. Brandenburger~\etal~\cite{DBLP:conf/dsn/BrandenburgerCL17/LCM} detect cloning and rollback attacks by sharing enclave state across distributed clients. Niu~\etal~\cite{DBLP:conf/ccs/NiuPZZ22/Narrator} track input–state transitions across distributed TEEs for consistency, while Briongos~\etal~\cite{DBLP:conf/acsac/BriongosKSW23/clonebuster} detect cloning via cache contention. These techniques incur significant overhead due to frequent state validation or cache probing. In contrast, \lib embeds cloning protection in TEE firmware, with only a one-time initialization cost.

\vspace{0.2cm}
\subheading{Firmware-supported enclave updates and migrations.}
To the best of our knowledge, no existing solution addresses secure enclave updates. Several works do study enclave migration for SGX~\cite{DBLP:conf/dsn/GuHXCZGL17/SGXmigrationCloud,DBLP:conf/dsn/AlderKPA18/SGXmigrationPersistentState,DBLP:conf/ucc/NakashimaK21/MigSGX,DBLP:conf/iscc/LiangZLL19/SGXmigrationContainers,DBLP:conf/eurosp/SorienteKLF19/replicaTEE} and SEV-SNP~\cite{DBLP:conf/ccs/BuhrenWS19/InsecureUntilProvenUpdated}, though only SEV-SNP implements migration support at the firmware level, as \lib does. However, SEV-SNP delegates certain security guarantees to the developer at the VM level, and while~\cite{DBLP:conf/ccs/BuhrenWS19/InsecureUntilProvenUpdated} provides the only publicly available description of SEV-SNP migration behavior, it omits details on integrity verification of transferred VM images, making it impossible to assess whether rollback protection is ensured (cf.~\Cref{tab:related_work}). Despite these limitations, firmware-level support offers clear advantages: it enables low-cost atomicity (S5) and reduces reliance on external components compared to application-layer solutions.

\vspace{0.2cm}
\subheading{Application-layer enclave updates and migrations.}
In SGX, version upgrades rely on signer-based sealing keys that can be derived across enclave versions. Since updates are handled entirely at the application layer without firmware enforcement, old and new versions may run concurrently (violating S5) while code integrity (S2), rollback protection (S4), and state continuity (S6) are left to enclave logic rather than enforced by the platform.
TrustZone-based TEEs embed version metadata directly in Trusted Application (TA) binaries. In OP-TEE, RPMB-backed storage can provide software rollback protection~\cite{DBLP:conf/uss/BuschMP24/spillTheTeA}, yet no standardized protocol exists for secure updates or migration, leaving state transfer and version management as the developer's responsibility.
Although several application-layer migration mechanisms have been proposed for SGX~\cite{DBLP:conf/dsn/GuHXCZGL17/SGXmigrationCloud,DBLP:conf/dsn/AlderKPA18/SGXmigrationPersistentState,DBLP:conf/ucc/NakashimaK21/MigSGX,DBLP:conf/iscc/LiangZLL19/SGXmigrationContainers,DBLP:conf/eurosp/SorienteKLF19/replicaTEE}, these approaches achieve at best only eventual atomicity and depend on additional management mechanisms such as leases and renewals. More importantly, as summarized in~\Cref{tab:related_work}, no prior work satisfies all six security properties we consider essential for TEE software lifecycle management. In some cases, insufficient detail in the described system model, threat model, or protocol made it impossible to assess certain properties; for instance,~\cite{DBLP:conf/iscc/LiangZLL19/SGXmigrationContainers} does not specify its authorization mechanisms.

\begin{table}[t]
    \centering
    \footnotesize
    \caption{Comparison of security properties achieved by prior migration work and \lib. ``$\star$'' indicates insufficient information to assess whether a property is satisfied.}
    \begin{tabular}{l c c c c c c c}
        \toprule
        \textbf{Migration} & & \multicolumn{6}{c}{\textbf{Security Property}} \\ \cmidrule{3-8}
        \textbf{solution} & \textbf{TEE} & \textbf{S1} & \textbf{S2} & \textbf{S3} & \textbf{S4} & \textbf{S5} & \textbf{S6}  \\
        \midrule

        \rowcolor{gainsboro}    
        \lib   
            & SGX     & \cmark  & \cmark & \cmark & \cmark  & \cmark & \cmark \\
        Alder \etal \cite{DBLP:conf/dsn/AlderKPA18/SGXmigrationPersistentState} 
            & SGX     & \cmark  & \cmark & \cmark & \cmark  & \xmark & \cmark \\
        Gu \etal \cite{DBLP:conf/dsn/GuHXCZGL17/SGXmigrationCloud} 
            & SGX     & \xmark  & \cmark & \cmark & \cmark  & \xmark & \cmark \\
        Liang \etal \cite{DBLP:conf/iscc/LiangZLL19/SGXmigrationContainers}   
            & SGX     & $\star$ & \cmark & \cmark & \cmark  & \xmark & \cmark \\
        Nakashima \etal \cite{DBLP:conf/ucc/NakashimaK21/MigSGX}   
            & SGX     & \xmark  & \cmark & \cmark & \cmark  & \xmark & \xmark \\
        Soriente \etal \cite{DBLP:conf/eurosp/SorienteKLF19/replicaTEE} 
            & SGX     & \cmark  & \cmark &  ---   & \cmark  & \cmark &  ---   \\
        Buhren \etal \cite{DBLP:conf/ccs/BuhrenWS19/InsecureUntilProvenUpdated}    
            & SEV-SNP & \xmark  & \cmark & \cmark & $\star$ & \xmark & \cmark \\ 
        \bottomrule
    \end{tabular}
    \label{tab:related_work}
\end{table}

\section{Conclusion}
In this work, we tackled a critical gap between static enclave execution and the dynamic operational demands of modern embedded systems---namely, the lack of essential mechanisms supporting the full enclave software lifecycle. To bridge this gap, we introduced \lib, a modular toolkit that augments RISC-V TEEs with secure enclave update, migration, state continuity, and trusted time functionality. While we instantiated \lib within the Keystone framework, it is designed for broad compatibility with other current and emerging RISC-V TEEs, including CURE and ACE.

Our evaluation demonstrates that \lib introduces negligible runtime and memory overhead while maintaining practical performance. Our prototype implementation achieves sub-second update latency, near-second migration times, microsecond-scale trusted time access, and almost zero-cost state continuity operations. This confirms that \lib satisfies the performance and reliability needs of contemporary IoT and automotive environments.

\section*{Acknowledgments}
This work is partly funded by the Deutsche Forschungsgemeinschaft (DFG, German Research Foundation) under Germany’s Excellence Strategy - EXC 2092 CASA - 390781972, and the European Union’s Horizon 2020 research and innovation program (REWIRE, Grant Agreement No. 101070627). 
Views and opinions expressed are, however, those of the authors only and do not necessarily reflect those of the European Union. Neither the European Union nor the granting authority can be held responsible for them.

\balance
\bibliographystyle{ACM-Reference-Format}
\bibliography{references}

@inproceedings{alder2023aboutTime,
    author = {Alder, Fritz and Scopelliti, Gianluca and Van Bulck, Jo and M\"{u}hlberg, Jan Tobias},
    title = {{About Time: On the Challenges of Temporal Guarantees in Untrusted Environments}},
    year = {2023},
    isbn = {9798400700873},
    publisher = {Association for Computing Machinery},
    address = {New York, NY, USA},
    url = {https://doi.org/10.1145/3578359.3593038},
    doi = {10.1145/3578359.3593038},
    booktitle = {Proceedings of the 6th Workshop on System Software for Trusted Execution},
    pages = {27–33},
    numpages = {7},
    location = {Rome, Italy},
    series = {SysTEX '23}
}

@inproceedings{DBLP:conf/isca/HoekstraLPPC13/sgx,
  author       = {Matthew Hoekstra and
                  Reshma Lal and
                  Pradeep Pappachan and
                  Vinay Phegade and
                  Juan del Cuvillo},
  editor       = {Ruby B. Lee and
                  Weidong Shi},
  title        = {{Using Innovative Instructions to Create Trustworthy Software Solutions}},
  booktitle    = {{HASP} 2013, The Second Workshop on Hardware and Architectural Support
                  for Security and Privacy, Tel-Aviv, Israel, June 23-24, 2013},
  pages        = {11},
  publisher    = {{ACM}},
  year         = {2013},
  url          = {https://doi.org/10.1145/2487726.2488370},
  doi          = {10.1145/2487726.2488370},
  bibsource    = {dblp computer science bibliography, https://dblp.org}
}

@techreport{whitepaper:sevSnp,
  author       = {{Advanced Micro Devices, Inc.}},
  title        = {{AMD SEV-SNP: Strengthening VM Isolation with Integrity Protection and More}},
  institution  = {{Advanced Micro Devices (AMD)}},
  year         = {2020},
  url          = {https://www.amd.com/content/dam/amd/en/documents/epyc-business-docs/white-papers/SEV-SNP-strengthening-vm-isolation-with-integrity-protection-and-more.pdf},
  note         = {White paper; accessed 2025-08-11}
}

@techreport{whitepaper:trustzone,
  author       = {{Arm Ltd.}},
  title        = {{ARM Security Technology: Building a Secure System using TrustZone Technology}},
  institution  = {{Arm Ltd.}},
  year         = {2009},
  url          = {https://documentation-service.arm.com/static/5f212796500e883ab8e74531},
  note         = {White paper (PRD29-GENC-009492C); 2025-08-11}
}

@techreport{whitepaper:tdx,
  author       = {{Intel Corporation}},
  title        = {{Intel Trust Domain Extensions (TDX) White Paper}},
  institution  = {{Intel Corporation}},
  year         = {2022},
  url          = {https://cdrdv2.intel.com/v1/dl/getContent/690419},
  note         = {White paper; accessed 2025-08-11}
}

@article{riscVmanual,
	title={{The RISC-V instruction set manual volume II: Privileged architecture version 1.7}},
	author={Waterman, Andrew and Lee, Yunsup and Avizienis, Rimas and Patterson, David A and Asanovic, Krste},
	journal={EECS Department, University of California, Berkeley, Tech. Rep. UCB/EECS-2015-49},
	year={2015},
    note={Accessed 2025-07-15}
}

@inproceedings{eurosys/LeeKSAS20/keystone,
  author       = {Dayeol Lee and
                  David Kohlbrenner and
                  Shweta Shinde and
                  Krste Asanovic and
                  Dawn Song},
  editor       = {Angelos Bilas and
                  Kostas Magoutis and
                  Evangelos P. Markatos and
                  Dejan Kostic and
                  Margo I. Seltzer},
  title        = {{Keystone: An Open Framework for Architecting Trusted Execution Environments}},
  booktitle    = {EuroSys '20: Fifteenth EuroSys Conference 2020, Heraklion, Greece,
                  April 27-30, 2020},
  pages        = {38:1--38:16},
  publisher    = {{ACM}},
  year         = {2020},
  url          = {https://doi.org/10.1145/3342195.3387532},
  doi          = {10.1145/3342195.3387532},
  bibsource    = {dblp computer science bibliography, https://dblp.org},
  note         = {DBLP:conf/eurosys/LeeKSAS20}
}

@inproceedings{uss/CostanLD16/sanctum,
  author       = {Victor Costan and
                  Ilia A. Lebedev and
                  Srinivas Devadas},
  editor       = {Thorsten Holz and
                  Stefan Savage},
  title        = {{Sanctum: Minimal Hardware Extensions for Strong Software Isolation}},
  booktitle    = {25th {USENIX} Security Symposium, {USENIX} Security 16, Austin, TX,
                  USA, August 10-12, 2016},
  pages        = {857--874},
  publisher    = {{USENIX} Association},
  year         = {2016},
  url          = {https://www.usenix.org/conference/usenixsecurity16/technical-sessions/presentation/costan},
  bibsource    = {dblp computer science bibliography, https://dblp.org}
}

@misc{arxiv/OzgaHLGSPJD25/ace_latest,
      title={{ACE: Confidential Computing for Embedded RISC-V Systems}}, 
      author={Wojciech Ozga and Guerney D. H. Hunt and Michael V. Le and Lennard Gäher and Avraham Shinnar and Elaine R. Palmer and Hani Jamjoom and Silvio Dragone},
      year={2025},
      eprint={2505.12995},
      archivePrefix={arXiv},
      primaryClass={cs.CR},
      url={https://arxiv.org/abs/2505.12995}, 
}

@inproceedings{hasp/Ozga23/ace,
  author       = {Wojciech Ozga},
  title        = {{Towards a Formally Verified Security Monitor for VM-based Confidential
                  Computing}},
  booktitle    = {Proceedings of the 12th International Workshop on Hardware and Architectural
                  Support for Security and Privacy, {HASP} 2023, Toronto, Canada, 29
                  October 2023},
  pages        = {73--81},
  publisher    = {{ACM}},
  year         = {2023},
  url          = {https://doi.org/10.1145/3623652.3623668},
  doi          = {10.1145/3623652.3623668},
  bibsource    = {dblp computer science bibliography, https://dblp.org}
}

@online{penglai_doc,
  author       = {{Penglai-Enclave}},
  title        = {{User manual for Penglai-TVM}},
  year         = {2026},
  url          = {https://penglai-doc.readthedocs.io/en/latest/Penglai-manual/User-Manual-TVM.html},
  note         = {Accessed 2025-07-15}
}

@inproceedings{DBLP:conf/uss/BahmaniBDJKSS21/cure,
  author       = {Raad Bahmani and
                  Ferdinand Brasser and
                  Ghada Dessouky and
                  Patrick Jauernig and
                  Matthias Klimmek and
                  Ahmad{-}Reza Sadeghi and
                  Emmanuel Stapf},
  editor       = {Michael D. Bailey and
                  Rachel Greenstadt},
  title        = {{CURE:} {A} {Security Architecture with CUstomizable and Resilient
                  Enclaves}},
  booktitle    = {30th {USENIX} Security Symposium, {USENIX} Security 2021, August 11-13,
                  2021},
  pages        = {1073--1090},
  publisher    = {{USENIX} Association},
  year         = {2021},
  url          = {https://www.usenix.org/conference/usenixsecurity21/presentation/bahmani},
  bibsource    = {dblp computer science bibliography, https://dblp.org}
}

@inproceedings{DBLP:conf/aspdac/PanPMZZY0LX025/Dep-TEE,
  author       = {Shangjie Pan and
                  Xuanyao Peng and
                  Zeyuan Man and
                  Xiquan Zhao and
                  Dongrong Zhang and
                  Bicheng Yang and
                  Dong Du and
                  Hang Lu and
                  Yubin Xia and
                  Xiaowei Li},
  editor       = {Yuichi Nakamura and
                  Yu Wang},
  title        = {{Dep-TEE: Decoupled Memory Protection for Secure and Scalable Inter-enclave
                  Communication on} {RISC-V}},
  booktitle    = {Proceedings of the 30th Asia and South Pacific Design Automation Conference,
                  {ASPDAC} 2025, Tokyo, Japan, January 20-23, 2025},
  pages        = {454--460},
  publisher    = {{ACM}},
  year         = {2025},
  url          = {https://doi.org/10.1145/3658617.3697763},
  doi          = {10.1145/3658617.3697763},
  bibsource    = {dblp computer science bibliography, https://dblp.org}
}

@inproceedings{DBLP:conf/uss/KuhneVS25/Dorami,
  author       = {Mark Kuhne and
                  Stavros Volos and
                  Shweta Shinde},
  editor       = {Lujo Bauer and
                  Giancarlo Pellegrino},
  title        = {{Dorami: Privilege Separating Security Monitor on {RISC-V} TEEs}},
  booktitle    = {34th {USENIX} Security Symposium, {USENIX} Security 2025, Seattle,
                  WA, USA, August 13-15, 2025},
  pages        = {1149--1166},
  publisher    = {{USENIX} Association},
  year         = {2025},
  url          = {https://www.usenix.org/conference/usenixsecurity25/presentation/kuhne},
  bibsource    = {dblp computer science bibliography, https://dblp.org}
}

@inproceedings{DBLP:conf/ccs/NiuPZZ22/Narrator,
  author       = {Jianyu Niu and
                  Wei Peng and
                  Xiaokuan Zhang and
                  Yinqian Zhang},
  editor       = {Heng Yin and
                  Angelos Stavrou and
                  Cas Cremers and
                  Elaine Shi},
  title        = {{NARRATOR:} {Secure and Practical State Continuity for Trusted Execution
                  in the Cloud}},
  booktitle    = {Proceedings of the 2022 {ACM} {SIGSAC} Conference on Computer and
                  Communications Security, {CCS} 2022, Los Angeles, CA, USA, November
                  7-11, 2022},
  pages        = {2385--2399},
  publisher    = {{ACM}},
  year         = {2022},
  url          = {https://doi.org/10.1145/3548606.3560620},
  doi          = {10.1145/3548606.3560620},
  bibsource    = {dblp computer science bibliography, https://dblp.org}
}

@inproceedings{DBLP:conf/uss/MateticAKDSGJC17/ROTE,
  author       = {Sinisa Matetic and
                  Mansoor Ahmed and
                  Kari Kostiainen and
                  Aritra Dhar and
                  David M. Sommer and
                  Arthur Gervais and
                  Ari Juels and
                  Srdjan Capkun},
  editor       = {Engin Kirda and
                  Thomas Ristenpart},
  title        = {{ROTE:} {Rollback Protection for Trusted Execution}},
  booktitle    = {26th {USENIX} Security Symposium, {USENIX} Security 2017, Vancouver,
                  BC, Canada, August 16-18, 2017},
  pages        = {1289--1306},
  publisher    = {{USENIX} Association},
  year         = {2017},
  url          = {https://www.usenix.org/conference/usenixsecurity17/technical-sessions/presentation/matetic},
  bibsource    = {dblp computer science bibliography, https://dblp.org}
}

@inproceedings{DBLP:conf/uss/BuschMP24/spillTheTeA,
  author       = {Marcel Busch and
                  Philipp Mao and
                  Mathias Payer},
  editor       = {Davide Balzarotti and
                  Wenyuan Xu},
  title        = {{Spill the TeA: An Empirical Study of Trusted Application Rollback
                  Prevention on Android Smartphones}},
  booktitle    = {33rd {USENIX} Security Symposium, {USENIX} Security 2024, Philadelphia,
                  PA, USA, August 14-16, 2024},
  publisher    = {{USENIX} Association},
  year         = {2024},
  url          = {https://www.usenix.org/conference/usenixsecurity24/presentation/busch-tea},
  bibsource    = {dblp computer science bibliography, https://dblp.org}
}

@inproceedings{DBLP:conf/acsac/BriongosKSW23/clonebuster,
  author       = {Samira Briongos and
                  Ghassan Karame and
                  Claudio Soriente and
                  Annika Wilde},
  title        = {{No Forking Way: Detecting Cloning Attacks on Intel {SGX} Applications}},
  booktitle    = {Annual Computer Security Applications Conference, {ACSAC} 2023, Austin,
                  TX, USA, December 4-8, 2023},
  pages        = {744--758},
  publisher    = {{ACM}},
  year         = {2023},
  url          = {https://doi.org/10.1145/3627106.3627187},
  doi          = {10.1145/3627106.3627187},
  bibsource    = {dblp computer science bibliography, https://dblp.org}
}

@inproceedings{DBLP:conf/eurosp/SorienteKLF19/replicaTEE,
  author       = {Claudio Soriente and
                  Ghassan Karame and
                  Wenting Li and
                  Sergey Fedorov},
  title        = {{ReplicaTEE: Enabling Seamless Replication of {SGX} Enclaves in the
                  Cloud}},
  booktitle    = {{IEEE} European Symposium on Security and Privacy, EuroS{\&}P
                  2019, Stockholm, Sweden, June 17-19, 2019},
  pages        = {158--171},
  publisher    = {{IEEE}},
  year         = {2019},
  url          = {https://doi.org/10.1109/EuroSP.2019.00021},
  doi          = {10.1109/EUROSP.2019.00021},
  bibsource    = {dblp computer science bibliography, https://dblp.org}
}

@inproceedings{DBLP:conf/dsn/BrandenburgerCL17/LCM,
  author       = {Marcus Brandenburger and
                  Christian Cachin and
                  Matthias Lorenz and
                  R{\"{u}}diger Kapitza},
  title        = {{Rollback and Forking Detection for Trusted Execution Environments
                  Using Lightweight Collective Memory}},
  booktitle    = {47th Annual {IEEE/IFIP} International Conference on Dependable Systems
                  and Networks, {DSN} 2017, Denver, CO, USA, June 26-29, 2017},
  pages        = {157--168},
  publisher    = {{IEEE} Computer Society},
  year         = {2017},
  url          = {https://doi.org/10.1109/DSN.2017.45},
  doi          = {10.1109/DSN.2017.45},
  bibsource    = {dblp computer science bibliography, https://dblp.org}
}

@inproceedings{DBLP:conf/ndss/WildeGSK25/OurNDSSForkingWay,
  author       = {Annika Wilde and
                  Tim Niklas Gruel and
                  Claudio Soriente and
                  Ghassan Karame},
  title        = {{The Forking Way: When TEEs Meet Consensus}},
  booktitle    = {32nd Annual Network and Distributed System Security Symposium, {NDSS}
                  2025, San Diego, California, USA, February 24-28, 2025},
  publisher    = {The Internet Society},
  year         = {2025},
  url          = {https://www.ndss-symposium.org/ndss-paper/the-forking-way-when-tees-meet-consensus/},
  bibsource    = {dblp computer science bibliography, https://dblp.org}
}

@misc{blog:embitel/automotive/TEEvsHSM,
    author      = {{Embitel}},
    title       = {{Choosing between TEE and HSM For Automotive Security Mechanism}},
    year        = {2025},
    url         = {https://www.embitel.com/blog/embedded-blog/choosing-between-tee-and-hsm-for-automotive-security-mechanism},
    note        = {Accessed 2025-08-06}
}

@misc{blog:embitel/automotive/TEEinECU,
    author      = {{Embitel}},
    title       = {{TEE Based ECU Security: A Foundation for Secure Automotive Systems}},
    year        = {2025},
    url         = {https://www.embitel.com/blog/embedded-blog/tee-based-ecu-security-for-secure-automotive-systems},
    note        = {Accessed 2025-08-06}
}

@article{article:armSecureAutomotiveRPMB,
	title      ={Architecting Secure Automotive Systems},
	author     ={Andrew Michael Jones},
	publisher  ={arm},
	year       ={2017},
    url        = {https://armkeil.blob.core.windows.net/developer/Files/pdf/white-paper/architecting-secure-automotive-systems.pdf?utm_source=chatgpt.com},
    note       ={Accessed 2025-11-12}
}

@inproceedings{DBLP:conf/dsn/GuHXCZGL17/SGXmigrationCloud,
  author       = {Jinyu Gu and
                  Zhichao Hua and
                  Yubin Xia and
                  Haibo Chen and
                  Binyu Zang and
                  Haibing Guan and
                  Jinming Li},
  title        = {{Secure Live Migration of {SGX} Enclaves on Untrusted Cloud}},
  booktitle    = {47th Annual {IEEE/IFIP} International Conference on Dependable Systems
                  and Networks, {DSN} 2017, Denver, CO, USA, June 26-29, 2017},
  pages        = {225--236},
  publisher    = {{IEEE} Computer Society},
  year         = {2017},
  url          = {https://doi.org/10.1109/DSN.2017.37},
  doi          = {10.1109/DSN.2017.37},
  bibsource    = {dblp computer science bibliography, https://dblp.org}
}

@inproceedings{DBLP:conf/dsn/AlderKPA18/SGXmigrationPersistentState,
  author       = {Fritz Alder and
                  Arseny Kurnikov and
                  Andrew Paverd and
                  N. Asokan},
  title        = {{Migrating {SGX} Enclaves with Persistent State}},
  booktitle    = {48th Annual {IEEE/IFIP} International Conference on Dependable Systems
                  and Networks, {DSN} 2018, Luxembourg City, Luxembourg, June 25-28,
                  2018},
  pages        = {195--206},
  publisher    = {{IEEE} Computer Society},
  year         = {2018},
  url          = {https://doi.org/10.1109/DSN.2018.00031},
  doi          = {10.1109/DSN.2018.00031},
  bibsource    = {dblp computer science bibliography, https://dblp.org}
}

@inproceedings{DBLP:conf/ucc/NakashimaK21/MigSGX,
  author       = {Kenji Nakashima and
                  Kenichi Kourai},
  editor       = {Ivona Brandic and
                  Rizos Sakellariou and
                  Josef Spillner},
  title        = {{MigSGX: a migration mechanism for containers including {SGX} applications}},
  booktitle    = {{UCC} '21: 2021 {IEEE/ACM} 14th International Conference on Utility
                  and Cloud Computing, Leicester, United Kingdom, December 6 - 9, 2021},
  pages        = {6:1--6:10},
  publisher    = {{ACM}},
  year         = {2021},
  url          = {https://doi.org/10.1145/3468737.3494088},
  doi          = {10.1145/3468737.3494088},
  bibsource    = {dblp computer science bibliography, https://dblp.org}
}

@inproceedings{DBLP:conf/iscc/LiangZLL19/SGXmigrationContainers,
  author       = {Hongliang Liang and
                  Qiong Zhang and
                  Mingyu Li and
                  Jianqiang Li},
  title        = {{Toward Migration of SGX-Enabled Containers}},
  booktitle    = {2019 {IEEE} Symposium on Computers and Communications, {ISCC} 2019,
                  Barcelona, Spain, June 29 - July 3, 2019},
  pages        = {1--6},
  publisher    = {{IEEE}},
  year         = {2019},
  url          = {https://doi.org/10.1109/ISCC47284.2019.8969644},
  doi          = {10.1109/ISCC47284.2019.8969644},
  bibsource    = {dblp computer science bibliography, https://dblp.org}
}

@inproceedings{DBLP:conf/ccs/BuhrenWS19/InsecureUntilProvenUpdated,
  author       = {Robert Buhren and
                  Christian Werling and
                  Jean{-}Pierre Seifert},
  editor       = {Lorenzo Cavallaro and
                  Johannes Kinder and
                  XiaoFeng Wang and
                  Jonathan Katz},
  title        = {{Insecure Until Proven Updated: Analyzing {AMD} SEV's Remote Attestation}},
  booktitle    = {Proceedings of the 2019 {ACM} {SIGSAC} Conference on Computer and
                  Communications Security, {CCS} 2019, London, UK, November 11-15, 2019},
  pages        = {1087--1099},
  publisher    = {{ACM}},
  year         = {2019},
  url          = {https://doi.org/10.1145/3319535.3354216},
  doi          = {10.1145/3319535.3354216},
  bibsource    = {dblp computer science bibliography, https://dblp.org}
}

@article{riscvForAutomotiveAI,
    author = {{The Linux Foundation}},
    title = {{RISC-V for Automotive AI Use Cases}},
    url = {https://riscv.org/wp-content/uploads/2025/04/RISC-V_AIOpportunitiesChallenges_042825.pdf},
    year = {2025},
    note={Accessed 2026-02-03}
}

@article{infineonRiscvForAutomotive,
    author = {{Infineon  Technologies AG}},
    title = {{Infineon brings RISC-V to the automotive industry and is first to announce an automotive RISC-V microcontroller family}},
    url = {https://www.infineon.com/press-release/2025/infatv202503-067},
    year = {2025},
    note={Accessed 2026-02-03}
}

@article{mobileyestats,
    author = {{Mobileye}},
    title = {EyeQ System-on-Chip},
    url = {https://www.mobileye.com/technology/eyeq-chip/},
    year = {2026},
    note={Accessed 2026-02-03}
}

@misc{autocryptTee,
  title        = {{AutoCrypt TEE – Trusted Execution Environment for Vehicles}},
  author       = {{AUTOCRYPT}},
  howpublished = {\url{https://autocrypt.io/products/tee/}},
  note         = {Accessed 2026-02-03},
  year         = {2026}
}

@misc{quintauris,
  title        = {{Quintauris – Advancing the Adoption of RISC-V}},
  author       = {{Quintauris}},
  howpublished = {\url{https://www.quintauris.com/}},
  note         = {Accessed 2026-02-03},
  year         = {2026}
}

@misc{adesso2025ota,
  author       = {Marko Borozan},
  title        = {{Over-the-air Updates: The Digital Revolution in the Automotive Industry}},
  year         = {2025},
  month        = {May 5},
  howpublished = {\url{https://www.adesso.de/en/news/blog/over-the-air-updates-the-digital-revolution-in-the-automotive-industry.jsp}}
}

@misc{globalplatformAutomotive,
    author       = {{GlobalPlatform}},
    title        = {{Automotive Initiative \& Task Force}},
    year         = {2026},
    url          = {https://globalplatform.org/task-forces/automotive-task-force/},
    note         = {Accessed 2026-02-05}
}

@misc{psacertifiedSnapdragonSA61xxP,
  author       = {{PSA Certified}},
    title        = {{Snapdragon\textregistered{} Automotive SA61xxP product family}},
    year         = {2026},
    url          = {https://products.psacertified.org/products/snapdragon-automotive-sa61xxp-product-family},
    note         = {Accessed 2026-02-05}
}

@misc{qualcomm2025snapdragon,
    author       = {{Qualcomm Technologies, Inc.}},
    title        = {{Qualcomm's Snapdragon Cockpit Platforms Power Smart, Intuitive AI-Driven Experiences in New All-Electric Mercedes-Benz Vehicles}},
    year         = {2025},
    month        = {Sep},
    url          = {https://www.qualcomm.com/news/releases/2025/09/qualcomm-s-snapdragon-cockpit-platforms-power-smart--intuitive-a}
}

@misc{qualcommMercedesBenz,
    author       = {{Qualcomm Technologies, Inc.}},
    title        = {{Qualcomm Automotive Partners – Mercedes-Benz}},
    year         = {2026},
    url          = {https://www.qualcomm.com/automotive/partners/mercedes-benz},
    note         = {Accessed 2026-02-05}
}

@misc{renesas2025toyota,
    author       = {{Renesas Electronics Corporation}},
    title        = {{Renesas Innovative Automotive Chips Drive Next-generation Multimedia System to Toyota and Lexus}},
    year         = {2021},
    url          = {https://www.renesas.com/en/about/newsroom/renesas-innovative-automotive-chips-drive-next-generation-multimedia-system-toyota-lexus},
    note         = {Accessed 2026-02-05}
}

@misc{trustonicRenesasConsortium,
    author       = {{Trustonic}},
    title        = {{Trustonic joins Renesas R-Car Consortium to increase the availability of cybersecurity solutions in the connected vehicle market}},
    year         = {2021},
    url          = {https://www.trustonic.com/news/trustonic-renesas-r-car-consortium-cybersecurity-solutions/},
    note         = {Accessed 2026-02-05}
}

@article{tlsHandshakeLatency,
    author = {Akavaram, Sravanthi},
    year = {2025},
    month = {05},
    pages = {870-880},
    title = {{Secure Communication Protocol for Distributed Environments}},
    volume = {7},
    journal = {Journal of Computer Science and Technology Studies},
    doi = {10.32996/jcsts.2025.7.4.101}
}

@online{our_artifact,
	author = {Annika Wilde and
	Samira Briongos and
	Claudio Soriente and
	Ghassan Karame},
	title = {{Keyfort}: {Repository} on {GitHub}},
	url = {https://github.com/RUB-InfSec/Keyfort},
	year         = {2026}
}

\end{document}